\documentstyle[aps]{revtex}
\begin{document}
\draft
\title{A Cloudy Bag Model for the \\
S-D Wave $\bar{K}-N$ System}
\author{Guangliang He and Rubin. H. Landau}
\address{Physics Department, Oregon State University, Corvallis, OR 97331}

\date{\today}

\maketitle

\begin{abstract}
The cloudy quark bag model is applied to the coupled ($\bar{K} N$,
$\Sigma \pi$, $\Lambda \pi$) system for $S-D$ partial waves.
Energy--dependent separable potentials are derived, as are new
numerical algorithms for solving the coupled Lippmann-Schwinger
equations.  The parameters of the model are fit to $K^-p$ scattering
and reaction cross sections, branching ratios, and mass spectra from
$K^-p\rightarrow \Sigma\pi\pi\pi, \Lambda\pi\pi\pi$. Within the
constraints of the model, the branching ratio and $\Sigma\pi$ mass
spectrum data are in conflict.  The $\Sigma_{P13}(1385)$ and
$\Lambda_{D03}(1520)$ resonances are found to be predominately
elementary bag states with considerable dressing for the
$\Sigma_{P13}$. The $\Lambda_{S01}(1405)$ appears as a complicated
composite systems arising from two poles. The model with certain
parameter sets does predict two sign changes in the real part of the
$\bar{K}N$ scattering amplitude near threshold, but they are not quite
at the correct energies to produce agreement with the sign of the
strong interaction shift of kaonic hydrogen.
\end{abstract}

\section{INTRODUCTION}

It is generally accepted that hadrons are composed of quarks and that
strong interactions among quarks will be explained by quantum
chromodynamics (QCD).  Unfortunately, QCD is still too complicated to
solve and so we construct phenomenological models which incorporate
confinement, asymptotic freedom, and the symmetries of QCD---yet still
permit analysis of experimental data and application to nuclear
few--body systems.  The MIT ``bag''\cite{chodos74.1} is a model for a
bubble in the QCD vacuum and imposes confinement by placing the quarks
in a finite bag of about $1$\,fm radius. If the model were interpreted
literally, this large a bag would shatter conventional nuclear physics
since nuclei would be filled more with quarks than nucleons.

Soon after its introduction it was realized that the bag model
violates chiral symmetry (conservation of quark helicity) --- the
second best symmetry ($\sim7\%$) of the strong
interactions\cite{pagels75}.  The violation arises from quark
reflections off the bag wall changing the quark's momentum but not its
spin. The symmetry is restored by coupling in a pion field to the
bag's surface\cite{chodos75,brown79,jaffe79}.  In the ``little brown
bag'' of Brown and Rho\cite{brown79}, an interior phase contains free,
massless quarks, while an exterior phase contains pions but no quarks.
The pressure of the pion field on the outside of the bag then
compresses the bag down to the very small $0.3\,$fm little brown bag.
Clearly with this small a bag there will be very few quark effects
expected inside of nuclei, and the conventional views of nuclei as
collections of nucleons exchanging mesons holds.

In the cloudy bag model (CBM) of Miller, Thomas, and
collaborators\cite{theberge82.2,thomas84}, the exterior pion field
gets quantized and penetrates the bag. While the CBM's Lagrangian is
non-linear, practical calculations are carried out as a perturbation
expansion with the zeroth order term yielding the MIT bag model. The
first tests of the CBM dealt with static properties of hadrons, such
as charge radii and magnetic moments, and were
successful\cite{thomas84,theberge83}. Latter tests dealt with dynamic
properties, such as pion-nucleon scattering, and required an extension
of the Lagrangian to incorporate volume in addition to surface
coupling\cite{thomas81.2}. Since the extension incorporated Weinberg's
effective $\pi N$ Lagrangian, it is not surprising that success was
found\cite{veit86,cooper87}.

In order to study the $KN$ and $\overline{K}N$ systems, strangeness
was incorporated into the CBM by extending the Lagrangian from SU(2)
to SU(3)\cite{veit85.1,veit85.2}. However, the complications of
thresholds and exotic $KN$ resonances led Veit et al.\cite{veit84} to
conclude that the model ``has trouble with resonances'' for both the
$\overline{K}N$ and $KN$ systems.  For example, they found a
$\overline{K}N$ scattering amplitude which appeared to have a
$\Lambda$ resonance signal approximately $5$ MeV below threshold
rather than the expected $27$ MeV below. Further analysis by Fink et
al.\cite{fink} indicated that the resonance signal arises from the
influence of the threshold cusp on a pole some $13$ MeV {\em above}
threshold.

The $\bar{K}N$ findings are interesting because they reflect on the
nature of the subthreshold $\Lambda$ resonance and possibly on the
puzzling experimental measurements of the $1S$ strong interaction
level shift in kaonic hydrogen\cite{davies79,izycki80,bird83}. The
hydrogen experiments indicate that the real part of the $K^{-}p$
scattering amplitude has either no or two sign changes right above
threshold\cite{fink,schnick87,tanaka,kumar}, whereas $K$ matrix
analyses and most potential models predict one sign change. The CBM
fits have two sign changes close to threshold, but on either side of
threshold.

It has not been clear to us what level of agreement to expect between
models of the $\bar{K}N$ coupled system and energy--dependent
scattering and reaction data. On the one hand the CBM assumption of a
rigid spherical bag and of point--like pion fields is possibly too
simple, yet on the other hand it may be that only more partial waves
are needed to get the energy dependence right. There is, after all,
experimental indication that significant $P$ wave enters at $184$
MeV/c\cite{ciborowski82}, and there is the $D$-wave resonance at $390$
MeV/c.  To describe additional data and to test the model further, we
have extended the $\bar{K}N$ cloudy bag model to include the $P$ and
$D$ partial waves. This extends the contact interaction to include
space--like terms in addition to time--like ones, and the general
interaction to include spin--orbit forces.  We have deduced effective
potentials for use in a Lippmann-Schwinger equation, and have fit the
potentials' parameters to data up to $520$ MeV/c.  We were aided in
our work by the similar kaon work of Veit et al.\cite{veit85.2};
similar, yet with the additional and stronger interactions available
to the antikaon making for more differences than similarities.

\section{The Cloudy Bag Lagrangian}

The MIT bag model's Lagrangian is\cite{chodos75,jaffe79}
\begin{equation}
{\cal L}_{MIT}(x) \ = \ (\frac{i}{2}\bar{q}
\stackrel{\leftrightarrow}{\not\!\partial} q - B) \, \theta_v
-\frac{1}{2}\bar{q}q \, \delta_s \label{eq:mit}
\end{equation}
Here $q$ is the quark field, $B$ is a universal constant called the
``bag pressure'', and for a nondeformable, static, spherical bag, the
bag ansatz is imposed by setting the volume step-function $\theta_v =
\theta(R-r)$ and the surface delta function $\delta_s =
\delta(R-r)$.  By demanding that the action $S = \int\!d^4x\,{\cal
L}(x)$ remains stationary under arbitrary changes in
$q,\bar{q},\theta_v,$ and $\delta_s$, we obtain the equations of
motion:
\begin{eqnarray}
i\not\!\partial q(x) &=& 0, \ \ \ \ \ \, (r < R),  \label{eq:dirac} \\
i\not\!nq(x) &=& q(x), \ \   (r = R), \label{eq:bc} \\
B &=& -\frac{1}{2}n\cdot\partial\left[\bar{q(x)}q(x)\right]_{r=R}
\label{eq:bp}
\end{eqnarray}
For the static case, the unit normal to the bag's surface is $n^{mu} =
(0,\hat{\bf r})$. Equation (\ref{eq:dirac}) is the Dirac equation
inside the bag, (\ref{eq:bc}) is the linear boundary condition of
confinement, and (\ref{eq:bp}) is the stability condition on the bag
pressure. We shall use solutions to these equations in constructing
the baryon-meson effective potentials.

The Lagrangian (\ref{eq:mit}) does not preserve chiral symmetry since
the surface term $\frac{1}{2}\bar{q}q\delta_s$ reverses the helicity
of the quark hitting the bag's surface.  To incorporate chiral
symmetry, a pion field $\vec{\phi}$ is coupled throughout the bag's
volume\cite{chodos75,brown79,jaffe79}:
\begin{equation}
{\cal L}_{CBM} =
\left(\frac{i}{2}\bar{q}\stackrel{\leftrightarrow}{\not\!\partial}q
-B\right)\theta_v
-\frac{1}{2}\bar{q}e^{i\vec\tau\cdot\vec\phi\gamma_5/f}q\delta_s
+\frac{1}{2}\left(D_\mu\vec\phi\right)^2
\label{eq:cbm.s}
\end{equation}
Here the arrow over $\vec{\phi}$ indicates its isovector nature,
$\vec{\tau}$ is a vector of Pauli matrices for $SU(2)$ (isospin), $f$
is the meson octet decay constant, and $D_\mu$ is the covariant
derivative:
\begin{eqnarray}
D_\mu\vec\phi & = & (\partial_\mu\phi)\hat\phi +
f\sin(\phi/f)\partial_\mu\hat\phi \label{eq:cd.pi}\\
\phi & = & \left(\vec\phi\cdot\vec\phi\right)^{1/2},  \ \ \
\hat\phi =  \vec\phi/\phi
\end{eqnarray}
While the Lagrangian (\ref{eq:cbm.s}) has no obvious predictions for
low-energy pion-nucleon scattering, Thomas\cite{thomas81.2}
transformed it to one containing a volume coupling of the covariant
derivative of the quark field:
\begin{eqnarray}
{\cal L}_{CBM} &=&
(\frac{i}{2}\bar{q}\stackrel{\leftrightarrow}{\not\!\!D}q - B)\theta_v
-\frac{1}{2}\bar{q}q\delta_s +
\frac{1}{2}\left(D_\mu\vec\phi\right)^2
+\frac{1}{2f}\bar{q}\gamma^\mu\gamma_5\vec\tau q
\cdot(D_\mu\vec\phi)\theta_v
\label{eq:cbm.v} \\
D_\mu q &=& \partial_\mu q -
i\left[\frac{\cos(\phi/f)-1}{2}\right]
\vec\tau\cdot\left(\hat\phi\times\partial_\mu\hat\phi\right)q
\label{eq:cd.q2}
\end{eqnarray}
This Lagrangian incorporates the
Weinberg-Tomozawa relation for zero--energy $S$-wave pion scattering.

\section{Application to $\bar{K}N$ Scattering}

\label{sec.apply}

To apply the Lagrangian (\ref{eq:cbm.v}) to $\bar{K}N$ scattering,
Veit et al.\cite{veit85.1,veit85.2,veit84} extended the internal
symmetry from flavor $SU(2) \times SU(2)$ to $SU(3) \times SU(3)$.
The quark field $q$ then becomes any member of the SU(3) triplet
$(u,d,s)$, the meson field $\vec\phi$ any member of the octet, the
$SU(2)$ Pauli matrices $\vec\tau$ are replaced by $SU(3)$ Gell-Mann
matrices $\vec\lambda$, and the cross product in the covariant
derivative (\ref{eq:cd.q2}) includes $SU(3)$ structure constants
$(\vec A \times \vec B)_c = \sum_{a,b}f_{abc}A_aB_b$.  Since $\vec\phi
= 0$ yields the MIT bag Lagrangian (\ref{eq:mit}), which was
successful for static baryon properties, Veit et al.\cite{veit85.1}
postulated that if the energy is low then small $\phi$ should be a
good approximation for scattering.  Accordingly, they expanded ${\cal
L}_{CBM}$ around $\phi = 0$ to obtain the linearized, volume--coupled,
SU(3)$\times$SU(3), CBM Lagrangian:
\begin{eqnarray}
{{\cal L}_{CBM}} &=& {\cal L}_{MIT} + {\cal L}_{M}
+{\cal L}_s+{\cal L}_c \label{pieces}\\
{\cal L}_{M} & = &  \frac{1}{2}(\partial_\mu\vec\phi)^2 \\
{\cal L}_s & = &
\frac{\theta_v}{2f}\bar{q} \, \gamma^\mu \, \gamma_5 \, \vec\lambda
q\cdot(\partial_\mu\vec{\phi}) \\
{\cal L}_c & = &
-\frac{\theta_v}{(2f)^2}\bar{q} \, \gamma^{\mu} \,
\vec\lambda\cdot(\vec\phi\times
\partial_{\mu}\vec\phi)q
\end{eqnarray}
where ${\cal L}_{MIT}$ (\ref{eq:mit}) describes the free bag, ${\cal
L}_{M}$ the free meson field, ${\cal L}_s$ the s-channel interaction,
and ${\cal L}_c$ the contact interaction.

In our application of the CBM to the $\bar{K}N$ system, we restrict
the energy to ($1250 \leq E_{cm} \leq 1550$)MeV, that is, to both
sides of the $\bar{K}N$ threshold at $1432$ MeV. For these energies we
consider the six, strangeness $-1$, baryon-meson ($BM$) channels which
couple strongly:
\begin{equation}
\label{coupled}
K^-p\rightarrow\left\{
\begin{array}{lrc}
K^-p & (``threshold'')\ 1432 &  \mbox{MeV}\\
\bar{K^0}n & -5 & \mbox{MeV} \\
\Sigma^-\pi^+   & +95  & \mbox{MeV} \\
\Sigma^0 \pi^0  & +104 & \mbox{MeV} \\
\Sigma^+ \pi^-  & +103 & \mbox{MeV} \\
\Lambda\pi^0 &  +181 & \mbox{MeV}
\end{array}\right.
\end{equation}
This system supports a number of resonances which appear in different
channels below and above threshold. For low energies the relevant ones
are $\Lambda_{LI\,2J = S01}(1405)$ and $\Sigma_{P13}(1385)$ below
threshold and $\Lambda_{D03}(1520)$ above.

The calculations to follow get rather complicated and so we have
placed many of the details in appendices. We will describe how we
extended the CBM up to $D$ waves and in each partial wave deduced
effective potentials to be used in the Lippmann--Schwinger equation
\begin{equation}
T_{\beta\alpha}({\bf k'}, \mu'; {\bf k}, \mu) =
V_{\beta\alpha}({\bf k'}, \mu'; {\bf k}, \mu)
+ \sum_{\gamma,\nu}\int\!d^3p\,
\frac{V_{\beta\gamma}({\bf k'}, \mu'; {\bf p}, \nu)
T_{\gamma\alpha}({\bf p}, \nu; {\bf k}, \mu)}{E + i\epsilon - E_\gamma(p)}
\label{eq:ls3d}
\end{equation}
Here the incident and final particles are in channels $\alpha$ and
$\beta$ (defined in table~\ref{tab.chnl}), have spin $\mu$ and $\mu'$,
and COM momentum ${\bf k}$ and ${\bf k'}$. The intermediate state is
channel $\gamma$ with spin $\nu$ and with a channel momentum
$k_{0\gamma}$ determined by setting the channel energy equal to the
system energy $E$:
\begin{eqnarray}
E_\gamma(k_{0\gamma}) &=& E \label{channelk} \\
E_{\gamma}(p) &=&= E_{1}(p) + E_{2}(p)
	= \sqrt{m_{1}^{2} + p^{2}} + \sqrt{m_{2}^{2} + p^{2}}
\label{channelE}
\end{eqnarray}
where the subscripts $1$ and $2$ refer to the specific particles in
channel $\gamma$.

The solution $T$ of the Lippmann--Schwinger equation (\ref{eq:ls3d})
automatically includes all iterations (ladder graphs) of the
potentials but ignores crossed meson lines --- which we assume to be
small\cite{theberge82.2}.  The derivation of the potentials
$V_{\beta\alpha} $ proceed via several steps. First in \S
\ref{A} we specify the coupled-channel Lippmann-Schwinger equation in
the partial wave basis. Next in \S
\ref{B} we convert the Lagrangian (\ref{eq:cbm.v})  and (\ref{eq:cd.q2})
into a Hamiltonian, and separate off pieces which
produce interactions.  Then we obtain the resonant potentials (\S
\ref{C}) and the contact potentials (\S
\ref{D}) from the Foch--space matrix elements:
\begin{equation}
V_{\beta\alpha} = \ \langle\beta| H_{c} | \alpha\rangle  +
	\sum_{B_{0}=S01,D03,P13} \,\langle\beta|H_{s} | B_{0}\rangle
\frac{1} {E-M_{B_{0}}}  \langle B_{0}| H_{s} | \alpha\rangle   \
 \label{pot}
\end{equation}
The first term in (\ref{pot}) describes direct scattering via an
elementary quark transition in the contact interaction as illustrated
in Fig.~\ref{fig.feyn}B. The $B_{0}$ sum in (\ref{pot}) is over the
processes, each second order in the $H_{s}$ Hamiltonian illustrated in
Fig.~\ref{fig.feyn}A, in which there are elementary resonance
intermediate states as shown in Fig.~\ref{fig.res}. Resonances with
bare masses $M_{S01}, M_{P13},$ and $ M_{D03}$ ---but no widths--- are
thus built into the potential (as opposed to generated by it).
However, when the potential gets used as the driving term in the
Lippmann--Schwinger equation (\ref{eq:ls3d}), these elementary
resonances get ``dressed'', that is, the resulting $T$ matrix contains
resonances at shifted masses and with finite widths.  Furthermore, we
treat the elementary masses as adjustable parameters and let the data
determine the best values. When the fitting is complete we then have
dressed resonance energies and widths as determined from the behavior
of the $T$ matrix, as well as best--fit values for the bare masses. As
we shall see, there is no guarantee that the bare masses and resonance
energies are close. In addition, the potential derived from the
contact interaction $H_{c}$ generates ``two--body'' or ``composite
resonances'' in $T$ which are not explicit in the potential and thus
not elementary.

\subsection{Partial Wave Lippmann-Schwinger Equation} \label{A}

We assume the standard spin $0 \times \frac{1}{2}$ partial wave
expansions for the $T$ matrix\cite{qmII}:
\begin{eqnarray}
\label{eq:partial}
T_{\beta\alpha}({\bf k'}, \mu'; {\bf k}, \mu) &  = &
\frac{1}{2\pi^2}\sum_l\left[(l+1)T_{\beta\alpha}^{l+}(k', k)
+lT_{\beta\alpha}^{l-}(k', k)\right]\delta_{\mu'\mu}P_l(x)
\nonumber \\
& & \hspace{2ex} + \frac{1}{2\pi^2}\sum_l\left[T_{\beta\alpha}^{l+}(k', k)
-T_{\beta\alpha}^{l-}(k', k)\right]
\langle\mu'|i \mbox{\boldmath$\sigma$}\cdot \hat{n} | \mu\rangle
P_l^\prime(x) \label{eq:tlj} \\*[1ex] T_{\beta\alpha}^{lj}(k',k) &=&
\frac{\pi}{2} \sum_{m, m, \mu \mu'}
\langle {l} {m'}; {\frac{1}{2}} {\mu'}| {l} {\frac{1}{2}}; {j} {M} \rangle
\langle {l}{m}; {\frac{1}{2}}{\mu} |{l} {\frac{1}{2}};{j}{M} \rangle
\nonumber \\
 & & \times \int\!d\Omega_{k'}d\Omega_k
Y_{lm'}^*(\hat k')Y_{lm}(\hat k) T_{\beta\alpha}({\bf k'}, \mu'; {\bf
k}, \mu) \label{eq:tproject}\\*[1ex]
T^{l\pm}_{\alpha\alpha}(k_{0\alpha},k_{0\alpha}) &=& \hspace{2ex}
\frac{-e^{i\delta_{l\pm}} \sin\delta_{l\pm}}{2
\mu_{\alpha} k_{0\alpha}}
\end{eqnarray}
where $x= \cos\theta_{kk'}$, $n = \hat k\times\hat k'$, and $T^{l\pm}$
is shorthand for $T^{j=l\pm\frac{1}{2}}$.  The partial wave expansion
for the potential $ V_{\beta\alpha}({\bf k'}, \mu'; {\bf k}, \mu)$ has
the same form as that for $T_{\beta\alpha}({\bf k'}, \mu'; {\bf k},
\mu)$ (\ref{eq:tlj}), so, for example, the projection for the
potential is:
\begin{eqnarray}
V_{\beta\alpha}^{lj}(k',k)  &=& \frac{\pi}{2}
 \sum_{m, m, \mu \mu'}
\langle {l}{m'};{\frac{1}{2}} {\mu'} | {l}{\frac{1}{2}};{j}{M} \rangle
\langle {l}{m};{\frac{1}{2}}{\mu} |{l} {\frac{1}{2}}; {j}{M} \rangle
\nonumber \\
 & & \times \int\!d\Omega_{k'}d\Omega_k Y_{lm'}^*(\hat k')Y_{lm}(\hat
k) V_{\beta\alpha}({\bf k'}, \mu'; {\bf k}, \mu) \label{eq:vproject}
\end{eqnarray}
After substitution of the partial wave expansions of $T$ and $V$ into
(\ref{eq:ls3d}), we obtain the coupled one--dimension integral
equations:
\begin{equation}
T_{\beta\alpha}^{lj}(k',k) = V_{\beta\alpha}^{lj}(k',k)
+ \frac{2}{\pi}\sum_\gamma\int_0^\infty\!dp\,
\frac{p^2 V_{\beta\gamma}^{lj}(k',p)T_{\gamma\alpha}^{lj}(p,k)}
{E + i\epsilon -E_\gamma(k)}
\label{eq:ls1d}
\end{equation}

We indicated in Eq. (\ref{coupled}) that even for zero kinetic energy
(the $1432$ MeV threshold), a $\bar{K}N$ pair couples to nearby
strangeness $-1$ charge channels. To use the derived potentials, we
evaluate their matrix elements between charge states which in turn are
expanded in the isospin states of those channels:
\begin{equation}
\label{eq:ibasis}
\begin{array}{lc}
 | K^-p\rangle  & =\\
| \bar{K}^0n\rangle  & = \\
| \Sigma^-\pi^+\rangle  & = \\
| \Sigma^0\pi^0\rangle  & = \\
| \Sigma^+\pi^-\rangle  & = \\
| \Lambda\pi^0\rangle  & =
\end{array}
\begin{array}{lllll}
\frac{1}{\sqrt{2}}| 0, 0\rangle
&+& \frac{1}{\sqrt{2}}| 1, 0\rangle  & & \\
\frac{-1}{\sqrt{2}}| 0, 0\rangle
&+& \frac{1}{\sqrt{2}}| 1, 0\rangle & & \\
\frac{1}{\sqrt{3}}| 0,0\rangle '
&-& \frac{1}{\sqrt{2}}| 1,0\rangle ' &+&
\frac{1}{\sqrt{6}}| 2,0\rangle \\
\frac{-1}{\sqrt{3}}| 0,0\rangle '
& & & + &\frac{\sqrt{2}}{\sqrt{3}}| 2,0\rangle \\
\frac{1}{\sqrt{3}}| 0,0\rangle '
&+& \frac{1}{\sqrt{2}}| 1,0\rangle ' & +&
\frac{1}{\sqrt{6}}| 2,0\rangle  \\
 & & \hspace{2.5ex} | 1, 0\rangle '' & & \\
\end{array}
\end{equation}
Here the prime and double prime distinguish states with the same
isospin but in different particle channels.  Because our initial
$\bar{K}N$ channel has no isospin--$2$ component, the $| 2,0\rangle $
state does not couple to it and so we eliminate $| 2,0\rangle $ and
renormalize the rest.  Using just two isospin channels for the three
$\Sigma\pi$ states reduces computing times by 50\%. In order to
include some isospin breaking, we do our calculations in the charge
basis with physical masses for the particles and use these relations
to transform between isospin and charge matrix elements.

To solve the Lippmann-Schwinger equation~(\ref{eq:ls1d}) we have
extended the Haftel-Tabakin technique\cite{haftel70} (details in
Appendix~\ref{app.num}).  First, rather than work with only the
principal part we also keep the delta function (imaginary)
part. Also, in handling coupled channels  we reorganize the
storage of matrices and instead of solving (\ref{eq:ls3d}) we
solve
\begin{equation}
[G^{-1} - V] [G T] = [V]
\end{equation}
We now can use Gaussian elimination on the symmetric $[G^{-1} - V]$ as
opposed to the inversion needed to solve (\ref{eq:ls3d}). Eliminating
inversion saves $\sim 30$\% in time and utilizing the symmetry of
$[V]$ and $[G^{-1} - V]$ saves $\sim 60$\%.

The Coulomb force is included exactly in all our bound state (kaonic
hydrogen) calculations, and of course it is crucial there. Even though
we have the theoretical tools to include the Coulomb force in
scattering, we do not include it in the total cross section
calculations because the issue is too confused by different
bubble--chamber experiments using different techniques for its removal
during analysis. We do include some charge--symmetry--breaking effects
by using physical masses, but otherwise we assume isospin is a good
symmetry at the field theory and potential level and only break it by
using physical masses in the energies for the kaon and lambda
channels.

\subsection{Hamiltonian} \label{B}

We deduce the Hamiltonian by following the canonical procedures
starting with the energy-momentum tensor
$T^{\mu\nu}$\cite{veit85.1,veit85.2}:
\begin{eqnarray}
T^{\mu\nu} &=&
\frac{\partial{\cal L}}{\partial(\partial_\mu q)}\partial^\nu q
+\partial^\nu\bar{q}
\frac{\partial{\cal L}}{\partial(\partial_\mu\bar{q})}
+\frac{\partial{\cal L}}{\partial(\partial_\mu\vec\phi)}
\partial^\nu\vec\phi
-g^{\mu\nu}{\cal L} \\
\hat H &=& \int\!d^3x\,T^{00}(x) = \hat H_{0} + \hat H_s + \hat H_c\\
\hat H_{0} &=& \hat H_{MIT} + \hat H_{M}\\
\hat H_s &=& -\int\!d^3x\,
\frac{\theta_v}{2f}\bar{q}\gamma^\mu\gamma_5\vec\lambda
q\partial_\mu\vec\phi  \label{eq:h.s.1} \\
\hat H_c &=&  H_{ct} + H_{cs} \nonumber \\
&=& \int\!d^3x\,
\frac{\theta_v}{4f^2}\bar{q}\gamma^{0}\vec\lambda
	\cdot(\vec\phi\times \partial_{0}\vec\phi)q  + \int\!d^3x\,
\frac{\theta_v}{4f^2}\bar{q}\sum_{i}^{3}\gamma^{i}\vec\lambda
	\cdot(\vec\phi\times \partial_{i}\vec\phi)q \label{eq:h.c.1}
\end{eqnarray}
Here $\hat H_{MIT}$ is the free bag Hamiltonian, $\hat H_{M}$ is the
free meson Hamiltonian, $\hat H_s$ is the s-channel interaction
(Fig.~\ref{fig.feyn}A), and ($\hat H_ct$, $\hat H_cs$) are the time
and space derivative parts of the contact or four--point interaction
(Fig.~\ref{fig.feyn}B).  We convert $\hat H_{0}$ into the Foch--space
form $H_{0}$, by projecting onto the space of colorless baryons and
expressing the meson field $\vec\phi$ in terms of annihilation and
creation operators\cite{thomas84,theberge83}:
\begin{eqnarray}
H_{0} &=& \sum_{B_0,B_0'}B_0^{\dagger} \
\langle B_0|\hat H_{0} | B_0'\rangle
\ B_0'        \label{eq:h.s.3} \\
	&=&\sum_{B_0} \sqrt{m_{B_0}^{2}+k^{2}} \, B_0^\dagger B_0
+ \sum_i \int\!d^3k \, \omega_k \, a_i^\dagger({\bf k})a_i({\bf k})
\label{fourier}
\end{eqnarray}
Here $B_0$ is the annihilation operator for a three--quark bag of type
$B_0$, $| B_0\rangle $ is the bare baryon state, $m_{B_0}$ is the MIT
bare bag mass, and $\omega_k = \sqrt{m_{M}^{2} + k^{2}}$ is the energy
of a free meson of momentum $k$.

\subsection{Resonance Potentials} \label{C}

As discussed in \S \ref{sec.apply} and illustrated by the Feynman
diagram in Fig.~\ref{fig.res}, for each meson--baryon channel
$\alpha$ we identify one of the three resonance states $B_{0}=
\Lambda_{S01}(1405), \Sigma_{P13}(1385)$ or $\Lambda_{D03}(1520)$ to
include. As indicated in equation (\ref{pot}), we build potentials
containing resonances within them by evaluating this Feynman diagram
with the proper initial and final state quark wavefunctions. In
Appendix~\ref{appa} we simplify the Hamiltonian $H_{s}$ which
generates these resonances, and in Appendices
\ref{appb}-\ref{appd} we evaluate the contribution to (\ref{pot})
from each of the three resonances.  After the partial wave projection
there results the potentials
\begin{eqnarray}
v_{\beta\alpha}^{(S01),lj}(k', k) & = &
\delta_{l0}\delta_{j\frac{1}{2}}
\frac{u_{ \Lambda_{S01} \beta}(k')u_{ \Lambda_{S01} \alpha}(k)}
{32f^{2} \pi\sqrt{\omega_{k'}\omega_{k}}}
\frac{\lambda^{\Lambda}_{\beta}\lambda^{\Lambda}_{\alpha} }
{E-M_{S01}} \nonumber \\ & & \times
\langle I_{B'} i_{B'}; I_{M'} i_{M'} |I I_{B'}
I_{M'}; 0 0 \rangle
\langle  I_{B} i_{B}; I_{M} i_{M}
| I_{B} I_{M};0 0 \rangle
\label{eq:v.a}\\*[1ex]
v_{\beta\alpha}^{(D03),lj}(k', k) & = & \
\delta_{l2}\delta_{j\frac{3}{2}}
\frac{u_{\Lambda_{D03} \beta}(k')u_{\Lambda_{D03}^\alpha}(k)}
{32 f^{2}\pi\sqrt{\omega_{k'}\omega_{k}}}
\frac{\lambda_\beta^\Lambda\lambda_\alpha^\Lambda}{E-M_{D03}}
\nonumber \\
& & \times\,
\langle {I_{B'}} {i_{B'}}; {I_{M'}} {i_{M'}}|{I_{B'}}
{I_{M'}}; {0} {0}\rangle
\langle {I_B} {i_B}; {I_M} {i_M}
| {I_B} {I_M}; {0} {0} \rangle
\label{eq:v.d03} \\*[1ex]
v_{\beta\alpha}^{(P13),lj}(k', k) &=&
\delta_{l1}\delta_{j\frac{3}{2}}
\frac{\left[N_sR\,j_0(\omega_sR)\right]^4}{24f^{2}\pi}
\frac{\lambda_\beta^\Sigma\lambda_\alpha^\Sigma}{E-M_{P13}}
\frac{j_1(k'R)j_1(kR)}{\sqrt{\omega_{k'}\omega_{k}}} \nonumber \\
& & \times
\langle {I_{B'}} {i_{B'}}; {I_{M'}} {i_{M'}}
| {I_{B'}} {I_{M'}}; {1} {0} \rangle
\langle {I_B} {i_B}; {I_M} {i_M}|{I_B}  {I_M}; {1} {0} \rangle
\label{eq:v.p13}
\end{eqnarray}
The vertex functions $u_{B_{0}\alpha}(k)$ for the $S$ and $D$
resonances are seen in the appendices to be proportional to integrals
over spherical Bessel functions (quark wavefunctions); for
$\Sigma_{P13}$ that substitution is already made in (\ref{eq:v.p13}).
The potentials (\ref{eq:v.a})-(\ref{eq:v.p13}) are clearly separable
and energy dependent. Further, each contains a pole at the real energy
$E=M_{B_{0}}$ arising from the elementary resonances we have
explicitly incorporated.  As discussed in \S \ref{sec.apply}, for any
value of $M_{B_{0}}$ the $T$ matrices generated by using these
potentials in the Lippmann--Schwinger equation will have poles at
complex energies whose real parts differ from the $M_{B_{0}}$'s, that
is, the resonances get dressed and acquire widths.  In addition,
because the values for $M_{B_{0}}$ are determined by the data fitting,
their best fit values will differ from those we use as input.

\subsection{Contact Potentials $v^{(ct)}$, $v^{(cs)}$} \label{D}

As shown in Fig.~\ref{fig.feyn}B, the contact interaction
$H_c$~(\ref{eq:h.c.1}) directly produces $BM\rightarrow B'M'$
scattering arising from an elementary quark transition\cite{veit85.1}.
Because $B'$ and $B$ belong to the baryon octet, the quarks in the
initial and final states must be in $1s$ states. We break the contact
potential into the two pieces $v^{(ct)}$ and $v^{(cs)}$ which are the
matrix elements of the time--derivative and space--derivative parts of
$H_c$:
\begin{equation}
v_{\beta\alpha}^{ct,cs}({\bf k'}, \mu', {\bf k}, \mu)
=  \langle\beta, {\bf k'}, \mu'|H_{ct,cs}| \alpha, {\bf k}, \mu \rangle
\end{equation}
For $S$-waves  the evaluation is straightforward:
\begin{eqnarray}
v^{(cs)}(\bf{k,k'}) &=& 0 \ \ \ (S \ \mbox{wave})\\
v_{\beta\alpha}^{(ct)}({\bf k'}, {\bf k}) &=&
\sum_I\lambda_{\beta\alpha}^{t,I}
\langle {I_{B'}} {i_{B'}}; {I_{M'}} {i_{M'}}|{I_{B'}}  {I_{M'}}; {I} {0}
\rangle
\langle {I_B} {i_B}; {I_M} {i_M}| {I_B} {I_M}; {I} {0} \rangle
\frac{u_{\beta\alpha}^{(ct)}(k', k)}{ 16f^{2} \pi^{3}\sqrt{\omega_{k'}
\omega_{k}}}
\label{eq:vc} \\
u_{\beta\alpha}^{(ct)}(k', k) &=&  N_s^2 (\omega_{k}+\omega_{k'})
\int_0^R\!dr\,r^2\left[j_0^2(\omega_sr)+j_1^2(\omega_sr)\right]
j_0(kr)j_0(k'r)
\end{eqnarray}
The coupling constants $\lambda_{\beta\alpha}^{t,I}$ follow from the
SU(6) wavefunction of the baryon octet\cite{gibson}, and are given in
Table~\ref{tab.couple}.

The increased spin and angular momentum coupling in $P$ and $D$ wave
makes the vertex functions more complicated. The time derivative part is:
\begin{equation}
v_{\beta\alpha}^{(ct)}({\bf k'},{\bf k}) =
\frac{-i}{64f^2 \pi^3 \sqrt{\omega_{k'}\omega_{k}} }
{}^{sf}\langle B'|\int\!d^3x\theta_v\bar{q}_{1s}f_{i'ij}
\lambda_j(\omega_{k}+\omega_{k'})
\gamma^0e^{i{\bf (k-k')\cdot r}}q_{1s}| B \rangle ^{sf}
\end{equation}
We evaluate it by substituting the $1s$ quark
wavefunction~(\ref{eq:q1s}), substituting the partial-wave expansion
of the plane wave, and integrating over solid angles to obtain:
\begin{eqnarray}
v_{\beta\alpha}^{(ct)}({\bf k'}\mu', {\bf k}\mu) & =  & \delta_{\mu'\mu}
\frac{\omega_{k}+\omega_{k'}}
{8 f^2 \pi^2\sqrt{\omega_{k}\omega_{k'}}}\,
 \sum_{lm}Y^*_{lm}(\hat k)Y_{lm}(\hat k')
N_s^2\int_0^R\!dr\,r^2[j_0^2(\omega_sr)
+j_1^2(\omega_sr)]j_l(kr)j_l(k'r)\nonumber \\
& & \times
\sum_I\lambda_{\beta\alpha}^{t,I}
\langle {I_{B'}} {i_{B'}};
{I_{M'}} {i_{M'}} | I_{B'} {I_{M'}};{I} {0} \rangle
\langle {I_B} {i_B};{I_M} {i_M}|{I_B} {I_M};{I}{0} \rangle
\end{eqnarray}
The partial wave matrix elements follow from the definition
(\ref{eq:vproject}):
\begin{eqnarray}
v_{\beta\alpha}^{(ct),lj}(k', k) & = &
\frac{N_s^2(\omega_{k}+\omega_{k}')}{ 16\pi f^2 \sqrt{\omega_{k'}\omega_{k}}}
\int_0^R\!dr\,r^2[j_0^2(\omega_sr)+j_1^2(\omega_sr)]j_l(kr)j_l(k'r)
\nonumber \\
& & \times\sum_I
\langle {I_{B'}} {i_{B'}}; {I_{M'}} {i_{M'}}
| {I_{B'}}  {I_{M'}};{I} {i}\rangle
\lambda_{\beta\alpha}^{t,I}
\langle {I_B} {i_B}; {I_M} {i_M}
| {I_B}  {I_M};{I} {i} \rangle
\label{eq:vct}
\end{eqnarray}

The derivation of the space derivative part
$v_{\beta\alpha}^{(cs)}({\bf k'}, {\bf k})$ is more complicated than
the time derivative part and we outline in Appendix \ref{space} how:
\begin{eqnarray}
v_{\beta\alpha}^{(cs),lj}(k', k) &=&
\frac{A_l^j N_s^2} {4\pi f^2 \sqrt{\omega_{k'}\omega_{k}}}
\int_0^R\!dr\,r j_0(\omega_sr)
j_1(\omega_sr) j_l(kr)j_l(k'r) \nonumber \\
 & & \times\sum_I
\langle {I_{B'}} {i_{B'}}; {I_{M'}} {i_{M'}}
| {I_{B'}}  {I_{M'}}; {I} {i} \rangle
\lambda_{\beta\alpha}^{s,I}
 \langle {I_B} {i_B}; {I_M} {i_M} | {I_B}   {I_M}; {I} {i} \rangle
\label{eq:pw.cs}\\*[1ex]
A^{j}_{l} &=&  2\sqrt{6l(l+1)(2l+1)}\,(-1)^{j+l+\frac{3}{2}}
\left\{ \protect{ \begin{array}{ccc}
\frac{1}{2} & \frac{1}{2} & 1 \\
l & l & j \end{array}  } \right\}
 = \left\{ \protect{ \begin{array}{rl}
2(l+1), & j = l-\frac{1}{2} \\
-2l,    & j = l+\frac{1}{2}
	       \end{array}  } \right.
\end{eqnarray}

The contact potentials (\ref{eq:vc}), (\ref{eq:vct}), and
(\ref{eq:pw.cs}) all are manifestly separable, energy ($\omega$)
dependent, and contain vertex factors arising from the quark
wavefunctions. They do not contain elementary resonances, although we
will find them to be strong enough to generate composite resonances.

\section{Comparison with Experiment}

We test the CBM by seeing how well we can reproduce five different
groups of data after adjusting seven parameters: the bag radius $R$,
the coupling constants $(f^{I=0}, f^{I=1}_{\bar{K}}, f^{I=1}_{\pi})$,
and the bare masses $(M_{S01}, M_{P13}, M_{D03})$.  In comparison to
potential models with adjustable coupling constants and ranges, the
CBM's $SU(3) \times SU(3)$ symmetry greatly reduces the number of
parameters---especially since the bare mass values mainly affect the
local position of resonance peaks.

\subsection{Scattering and Reaction Cross Sections}

The first and largest data group is 300 measurements of two--body
scattering and reactions cross sections for ($70  \leq K_{lab}
\leq 513$) MeV/c $\equiv$  ($1435 \leq E_{cm} \leq 1567$ MeV)
\cite{ciborowski82,watson63,sakitt65,kim,kittel66,mast,bangerter81,evans83}.
These data, shown in Figs.~\ref{fig.fit1+2}-\ref{fig.fitadd}, are
predominantly from bubble chambers and contain rather large
statistical and systematic errors.  The extension of the cloudy bag
model to higher partial waves is important for understanding these
data since even for a $K^-$ momentum as low as $150$ MeV/c, $P$-wave
contributions to $K^-p\rightarrow\Sigma^\mp\pi^\pm$ have been
reported\cite{ciborowski82}, and at $390$ MeV/c there is the
$\Lambda(1520)$ $D$-wave resonance (the peak evident in the figures
near $400$ MeV/c)\cite{watson63}.

The cross sections are related to the computed $T$-matrix elements
(\ref{eq:ls1d}) by
\begin{eqnarray}
\frac{d\sigma_{\beta\alpha}}{d\Omega} &=& 16\pi^4\mu_{\beta}\mu_{\alpha}
\frac{k'}{k} \, \overline{\sum_{\mu',\mu}}\left|T_{\beta\alpha}
({\bf k'}\mu', {\bf k}\mu)\right|^2 \\
\sigma_{\beta\alpha} &=& \int\!d\Omega\,\frac{d\sigma}{d\Omega}
= 16\pi\mu_\beta\mu_\alpha\frac{k'}{k}
\sum_{l=0}^\infty\left[(l+1)\left|T_{\beta\alpha}^{l+}(k',k)\right|^2
+l\left|T_{\beta\alpha}^{l-}(k',k)\right|^2
\right]
\end{eqnarray}
Here the bar indicates a sum over final and average over initial spin
states, and the $\mu_{\alpha}$'s are relativistic reduced channel
masses defined in Eq.~(\ref{eq:reduced}).  For the channels we use
the integrated cross sections are:
\begin{eqnarray}
\sigma(K^-p\rightarrow K^-p) & = &
16\pi\mu_1^2\sum_{l=0}^2
\left[(l+1)\left|T_{11}^{l+}(k', k)\right|^2
+l\left|T_{11}^{l-}(k', k)\right|^2\right] \\
\sigma(K^-p\rightarrow \bar{K^0}n) & = &
16\pi\mu_1\mu_2\frac{k'}{k}
\times\sum_{l=0}^2\left[(l+1)\left|T_{21}^{l+}(k', k)\right|^2
+l\left|T_{21}^{l-}(k', k)\right|^2\right]\\
\sigma(K^-p \rightarrow \Sigma^+\pi^-)
 & = & 16\pi\mu_3\mu_1\frac{k'}{k}\sum_{l=0}^2
\left[(l+1)\left|\frac{1}{\sqrt{3}}T_{31}^{l+}(k', k)
                +\frac{1}{\sqrt{2}}T_{41}^{l+}(k', k)
\right|^2\right. \nonumber \\
& & \left. \hspace{3ex} +l\left|\frac{1}{\sqrt{3}}T_{31}^{l-}(k', k)
       +\frac{1}{\sqrt{2}}T_{41}^{l-}(k', k)\right|^2\right] \\
\sigma(K^-p\rightarrow \Sigma^0\pi^0)
 & = & 16\pi\mu_3\mu_1\frac{k'}{k}\sum_{l=0}^2
\left[(l+1)\left|\frac{1}{\sqrt{3}}T_{31}^{l+}(k', k)\right|^2
+ l\left|\frac{1}{\sqrt{3}}T_{31}^{l-}(k', k)\right|^2\right] \\
\sigma(K^-p\rightarrow \Sigma^-\pi^+)
 & = & 16\pi\mu_3\mu_1\frac{k'}{k}\sum_{l=0}^2
\left[(l+1)\left|\frac{1}{\sqrt{3}}T_{31}^{l+}(k', k)
-\frac{1}{\sqrt{2}}T_{41}^{l+}(k', k)\right|^2\right. \nonumber \\
& & \hspace{3ex} \left.+ l\left|\frac{1}{\sqrt{3}}T_{31}^{l-}(k', k)
-\frac{1}{\sqrt{2}}T_{41}^{l-}(k', k)\right|^2\right]\\
\sigma(K^-p\rightarrow \pi^0\Lambda) &=&
16\pi\mu_5\mu_1\frac{k'}{k}\sum_{l=0}^2
\left[(l+1)\left|T_{51}^{l+}(k', k)\right|^2
+l\left|T_{51}^{l-}(k', k)\right|^2\right]
\end{eqnarray}

\subsection{The $\Lambda(1405)$ Resonance}

The second data group we examine is seven values of the
$\Sigma^+\pi^-$ mass spectrum determined from the $K^-p \rightarrow
\Sigma^+\pi^-\pi^+\pi^-$ reaction at 4.2~GeV/c kaon lab
momentum by Hemingway\cite{hemingway85}. These data, shown in
Figs.~\ref{fig.fit1+2}-\ref{fig.fitadd}, have a strong $\Lambda(1405)$
resonance signal and are a major constraint on the model's parameters.
Earlier $\Lambda(1405)$ data also exist, but Hemingway's are cleaner
since they were obtained through the three--step process $K^{-}p
\rightarrow
\Sigma^{+}\pi^{-}$, $\Sigma^{+} \rightarrow
\Lambda(1405) \pi^{+}$, $\Lambda(1405) \rightarrow \Sigma^{0}\pi^{0}$.
We calculate the $\Sigma\pi$ mass spectrum with a
Watson model\cite{watson63,dalitz91} which assumes an $S$--wave
resonance dominates the final state interaction. The number of events
per unit energy interval is accordingly:
\begin{equation}
\frac{dN}{dE}  \ \ \propto \ \ k\,\sigma_{\Sigma\pi\rightarrow\Sigma\pi}
	    \ \    \propto \ \
k\mu_{3}^2\left|T_{33}^0(k, k)\right|^2
\end{equation}
We fix the normalization by setting the area under the
theoretical mass spectrum equal to the total number of events.
Because this model is so simple, we do not expect a detailed
reproduction of the spectrum.

\subsection{The $\Sigma(1385)$ Resonance}

The $\Sigma(1385)$ is seen as a $P$-wave $\pi^0\Lambda$ resonance
below the $\bar{K}N$ threshold.  Aguilar-Benitez and
Salicio\cite{aguilar81} have detected it in the mass spectrum from
$K^-p\rightarrow\Lambda\pi^0\pi^+\pi^-$ at 4.2~GeV/c.  Our calculation
of that $\Lambda\pi$ spectrum is similar to the $\Sigma\pi$
calculation, only now with the Watson final state interaction model
adapted to a $P$-wave resonance:
\begin{equation}
\frac{dN}{dE}  \ \ \propto  \ \
              \frac{\sigma_{\Lambda\pi\rightarrow\Lambda\pi}}{k}
	    \ \    \propto \ \
\frac{\mu_{5}^2}{k}\left|T_{55}^{1+}(k, k)\right|^2
\end{equation}
Because the resonance rests upon a large background, and because we
have no model for such a background, we extract the resonance piece
after fitting the entire spectrum as a polynomial background plus a
Breit--Wigner resonance\cite{aguilar81}:
\begin{eqnarray}
\frac{dN}{dM} &=& a_0 + a_1\left(\frac{M}{M_0}\right)
+a_2\left(\frac{M}{M_0}\right)^2
+ \frac{bM_0^2\Gamma_0^2}{(M-M_0)^2+M_0^2\Gamma^2}\\*[1ex]
	(M, \Gamma) &=& (1384, 34.8) \mbox{MeV},
\ \ \  a_{0,1,2}  = (-77554, 15303, -7402) \mbox{MeV}, \ \
	b = 158.7 \mbox{MeV}^{-4}
\end{eqnarray}
The resulting resonance is shown in
Fig.~\ref{fig.fit3+4} as the peak at $1385$ MeV.

\subsection{The $K^-p$ Threshold Branching Ratios}

The third group of data we examine is the branching ratios of $K^{-}p$
reaction rates at threshold:
\begin{equation}
\gamma =  \frac{K^-p\rightarrow\Sigma^-\pi^+}
{K^-p\rightarrow\Sigma^+\pi^-}, \ \
R_c  =  \frac{K^-p\rightarrow\mbox{charged}}
{K^-p\rightarrow\mbox{all}}, \ \
R_n  =  \frac{K^-p\rightarrow\pi^0\Lambda}
{K^-p\rightarrow\mbox{neutral}}
\end{equation}
The measured ratios\cite{humphrey62,tovee71,nowak78,goossens80} are
shown in Fig.~\ref{fig.br}.  The attraction of ratios is that they are
not affected by uncertainties in target size and beam normalization,
and so their uncertainties are smaller than those of cross sections.
The difficulty is that calculations of rates into charge channels at
zero energy are sensitive to Coulomb and charge--symmetry--breaking
effects --- not all of which are included in our model --- and so we
cannot be sure of the level of agreement to expect.

\subsection{Data Fitting}

The parameters of our model determined by various fits to these data
groups are given in Table~\ref{tab.params}. The comparisons with the
cross section and mass data are in
Figs.\ref{fig.fit1+2}-\ref{fig.fitadd}, and the experimental and
theoretical branching ratios are in Fig.~\ref{fig.br}.  The data
groups have large differences in the number of data points, the type
of data, and the accuracy of the error estimates.  Accordingly, it was
not clear to us how properly to perform a best fit to all groups
simultaneously.  Consequently we have tried a variety of fitting
procedures (indicated in Table~\ref{tab.fits}) in which different
groups of data were systematically included and excluded, and with
different choices for the $\chi^{2}$ weights for each.

In Fig.~\ref{fig.fit1+2} we see fit 1. It employs only $S$-waves and
has its parameters adjusted to fit the low energy ($p \leq 250$ MeV/c,
$E \leq 1470$ MeV) two--body scattering and reaction cross sections.
This fit provides a good base for comparison with the results of other
groups and with our latter results in which higher partial waves are
included.  Fit 2 is also an $S$-wave fit to low energy data, but
extends fit 1 by including the $\Sigma\pi$ mass spectrum with its
$\Lambda_{S01}(1405)$ resonance.  The fit labeled TRIUMF (A) uses the
CBM parameter set A given by Veit et al.\cite{veit85.1} and uses the
same criteria as our fit 2. Fit 2 and TRIUMF produce somewhat
different parameters (row 1 {\em vs} 3 in Table~\ref{tab.params}) due
to differences in procedures and data.

We see in Fig.~\ref{fig.fit1+2} how much better an agreement fit 2
provides with the $\Sigma\pi$ mass spectrum than fit 1, and in
Table~\ref{tab.params} the large extent to which inclusion of the
$\Sigma\pi$ spectrum affects the model's parameters. In examining the
predictions for the branching ratios in the top part of
Fig.~\ref{fig.br}, we notice that fit 1 provides close agreement with
$R_{n}$ and $R_{c}$, and fairly close agreement with $\gamma$.
However, fit 2, which also included the $\Sigma\pi$ mass spectrum in
its fitting, provides noticeably worse agreement with $\gamma$. For
contrast, in the bottom part of Fig.~\ref{fig.br} we present the
branching ratios calculated with the Schnick--Landau potential
model\cite{schnick87}.  We see that the updated parameters determined
by Tanaka and Suzuki\cite{tanaka} provide excellent agreement with the
ratios.

In Fig.~\ref{fig.fit1+2} we notice that all three $S$--wave fits fail
to obtain agreement with the data at higher energies, and in
particular, they fail to reproduce the $\Lambda_{D03}(1520)$ resonance
signal at $400$ MeV/c. This clearly shows the need for higher partial
waves and additional resonances.

Our next two fits are shown in Fig.~\ref{fig.fit3+4}. Fit 3 uses $S$,
$P$, and $D$ waves, examines data at higher momenta, and includes the
three mass spectra for the $\Lambda_{S01}(1405)$,
$\Sigma_{P13}(1385)$, and $\Lambda_{D03}(1520)$ resonances. To
determine the influence of the branching ratio data and to deemphasize
the $\Sigma\pi$ spectrum, in fit 4 we do not fit to branching ratios
and manually (the ``m'' in Table ~\ref{tab.fits}) set the $\chi^{2}$
weight factors to $(w_{scattering}, w_{\Sigma\pi}, w_{\Lambda\pi}) =
(6, 1, 6)$.

It is interesting to note in Fig.~\ref{fig.fit3+4} that fits 3 and 4,
which include $S-D$ waves, do not provide as good agreement with the
$K^{-}p \rightarrow (K^{-}p, \bar{K}^{0}n, \Lambda^{0}\pi^{0})$
elastic cross sections as the $S$--wave fits 1 and 2. What is clear
from the figure, however, is that fits 3 and 4 provide nearly perfect
agreement with the $K^{-}p \rightarrow \Sigma\pi$ cross sections;
apparently the large number of high energy, small error, $\Sigma\pi$
data points dominate these fits. This analysis is confirmed by
confirming that the fit with the high energy data have a lower
$\chi^{2}$ per degree of freedom.

We again see in the $\Sigma\pi$ mass spectrum of Fig.~\ref{fig.fit3+4}
that fit 3 (which was adjusted to fit the branching ratios) does not
produce a good $\Sigma\pi$ mass spectrum.  Yet by releasing the
branching ratio constraint we obtain fit 4---which agrees with the
spectra quite well.  We also see that both fits do provide fairly good
agreement with the $\pi\Lambda$ mass spectrum. This spectrum shows the
sub $\bar{K}N$--threshold $\Sigma_{P13}(1385)$ resonance and
essentially determines the value for $M_{P13}$.

It can be argued that because so many assumptions go into applying the
Watson final state interaction model to fit the various mass spectra,
we should not expect good agreement with those spectra, and so should
not use the spectra in data fitting. For these reasons, in fits n7 and
c1 we used all partial waves but fit only the two--body scattering and
reaction data; n7 has weights determined by the experimental errors,
c1 has equal weights so the small $\Sigma\pi$, error bars at high
energy would have less influence.  Again we note that without the
$\Sigma\pi$ mass spectrum constraint, very good agreement with the
branching ratios is obtained (fit n7 in Fig.~\ref{fig.br}) even though
the ratio data were not included in the fit. Yet if the $\Sigma\pi$
mass spectra must be fit, then fit 4 is our best fit even though we
see in Fig.~\ref{fig.br} that it does not fit the branching ratios
well.

Because the $\pi\Lambda$ mass spectrum determines the value for
$M_{P13}$, and because this resonance does not significantly affect
the cross section data above the $\bar{K}N$ threshold, not fitting
this spectrum means the value for $M_{P13}$ were not determined in
fits n7 and c1.  This explains the negative values they have in
Table~\ref{tab.params}. The other resonance masses do affect the cross
section data above the $\bar{K}N$ threshold, and reasonable values
were obtained for them.

\section{The Scattering Amplitudes}

Now that the model parameters are determined, for each parameter set
we examine the scattering amplitudes as a function of energy.  We
shall see that these amplitudes often have such a rapid energy
dependence, especially near the $\bar{K}N$ threshold, that an
examination of the full energy dependence is more significant that
just the threshold values (scattering lengths). Values at that one
energy are too sensitive to details such as channel mass
values\cite{cheng}.

We start in Fig.~\ref{fig.TPD} with the scattering amplitudes for
$\bar{K}N \rightarrow \bar{K}N$ in the $D_{03}$ channel and for the
$\pi\Lambda \rightarrow \pi\Lambda$ in $P_{13}$.  As we move down in
energy we see vertical bars which signal the $\bar{K}N$ and
$\Sigma\pi$ thresholds (the $\pi\Lambda$ threshold at $1252$ MeV is
not visible).  We also see cusps in $f$ at channel openings and that
$\mbox{Im } f$ --- but not $\mbox{Re} f$ --- vanishes when all
channels are closed.  The $\bar{K}N$ resonance $\Lambda_{D03}(1520)$
and the $\pi\Lambda$ resonance $\Sigma_{P13}(1385)$ are clearly
signaled by $\mbox{Re} f$ changing sign and $\mbox{Im } f$ peaking at
approximately the energy corresponding to the resonance mass. The bare
versions of these resonances were built into our effective potentials
via (\ref{pot}), yet this does not guarantee that they will remain
after their dressing by the Lippmann--Schwinger equation and the
adjustment of parameters in the data fitting.

In Fig.~\ref{fig.TS} we give the $I=0$, $S$-wave $\bar{K}N$ scattering
amplitude for fits 1-4 and TRIUMF-A (since fit 2 is an updated
TRIUMF-A, the two are very similar).  We see that fit 1's $\mbox{Re}
f$ changes sign once below the $\bar{K}N$ threshold (right--most arrow
in Fig.~\ref{fig.TS}) and once again slightly above the $\Sigma\pi$
threshold (left--most arrow).  All the fits' $\mbox{Re} f$'s pass
through or very close to zero above and below the $\bar{K}N$
threshold, but only fit 1's $\mbox{Re} f$' has an extra sign change
far below. This is fascinating, for the ability of the model to
reproduce the experimental $1S$ strong interaction shift of kaonic
hydrogen depends on the magnitude and sign of $\mbox{Re} f$ at
threshold, and clearly there are several sign changes here.

The imaginary parts of $f$ (lower part of figure) for fits 1-3 show a
shoulder right below the $\bar{K}N$ threshold at $\sim 1440$ MeV,
while fit 4 shows a peak at $\sim 1425$ MeV. None of these behaviors
for $\mbox{Re} f$ and $\mbox{Im } f$ are the classical resonance
signal expected at $1405$ MeV for the $\Lambda_{S01}(1405)$. Although
there appears to be a resonance signal somehow mixed into the
$\bar{K}N$ threshold behavior.  Fits 1 and 3 do show what appears to
be distinctive resonance signals in both $\mbox{Re} f$ and $\mbox{Im }
f$, but they appear way down in energy near the $\Sigma\pi$ threshold
of $1325$ MeV! Clearly, since Table~\ref{tab.params} shows the fitted
$M_{S01} \geq 1550$ MeV, the cloudy bag model does not give the
$\Lambda_{S01}(1405)$ as a simple bag state of mass $\simeq 1405$ MeV.

In order to unravel some of the mysterious resonance behavior of the
$S_{01}$ $\bar{K}N$ scattering amplitude, we have solved for the
complex energies at which the scattering amplitudes have poles.  If
the pole is on the second (unphysical) energy sheet near the real
positive axis, it will produce an observable, narrow resonance along
the positive real energy axis with a width related to the imaginary
part of the complex energy by
\begin{equation}
\Gamma \simeq  -2 \mbox{Im } E
\end{equation}
Other singularities or channel openings can interfere with this
picture.  If the pole is far away from the real energy axis or some
other singularity gets in the way, its experimental and dynamical
significance is less.  In Appendix~\ref{app.num} we show that the
formal solution of the Lippmann--Schwinger equation implies that the
pole positions are solutions of the equation
\begin{equation}
\det(1-V_{E}G_{E}) = 0 \label{pole}
\end{equation}
We find numerically the complex energies which solve (\ref{pole}).

In Fig.~\ref{fig.3D} we have plots of the imaginary parts of the
$\bar{K}N$ $S$ wave scattering amplitudes of fits 2 (upper part of
figure) and 3 (lower part of figure) as functions of complex energy.
As found in the explicit solution of (\ref{pole}), and evident in the
figure, there are two poles in the $S01$ channel at the complex energy
listed in Table~\ref{tab.pole}.  For fit 2 we note that the high
energy pole is actually $17$ MeV {\em above} the $\bar{K}N$ threshold
(the tick mark to the right of $1400$ MeV in Fig.~\ref{fig.3D}), and
$31$ MeV along the negative imaginary axis.  Fit 2's low energy pole
is seen to be $80$ MeV below the $\bar{K}N$ threshold and far from the
real energy axis. In contrast, fit 1 and fit 3 (in the lower part of
the figure) have low energy poles very close to the real energy axis.
They are the causes of the striking energy dependences near the
$\Sigma\pi$ threshold seen in Fig.~\ref{fig.TS}.

In contrast to these CBM fits, we note in Table~\ref{tab.pole} that
the potential models of Alberg et al.\cite{alberg76} and Schnick and
Landau\cite{schnick87} have only one $S$--wave pole, and it is closer
to the tabulated $\Lambda(1405)$ energy. Apparently in the potential
model, the $\Lambda(1405)$ is a composite resonance with a single pole
close to the tabulated energy, while in the cloudy bag model, the
$\Lambda(1405)$ is not elementary and not simple, a conclusion drawn
previously from a number of viewpoints
\cite{veit85.1,fink,dalitz91,alberg76,umino89}.

The $T$ matrices' pole positions for the $P$ and $D$ wave amplitudes
are also listed in Table~\ref{tab.pole}. These positions correspond to
a $\Sigma_{P13}(1395)$ dressed mass of $1385$ MeV and width of $24$
MeV, and a $\Lambda_{D03}(1520)$ dressed mass of $1514$ MeV and width
of $12$ MeV.  The values for the best--fit bare masses are $M_{P13}
\simeq 1419$MeV and $M_{D03} \simeq 1552$ MeV.  This means that in
{\em both} cases the renormalization by the contact interaction and
higher order scatterings shift the masses downwards in energy by $31$
MeV!. The tabulated\cite{pdg90} masses and widths of $(1383.7\pm1.0$,
$36\pm5)$ MeV and $(1519.5\pm1.0$, $15.6\pm1.0)$ MeV are in excellent
agreement with the pole positions --- especially since the two are
expected to differ somewhat when the pole is not close to the real
energy axis. The tabulated half widths of the $\Sigma_{P13}(1395)$ and
$\Lambda_{D03}(1520)$ are not in as close agreement with the imaginary
parts of the pole energies ($18\pm3\ vs\ 12$, $8\pm1.0\ vs \ 6$), but
this is expected since the widths arise completely from
renormalization, and this is a broad and nonsymmetric resonance.

\subsection{Kaonic Hydrogen}

There are three measurements of the width $\Gamma$ and strong
interaction shift relative to Bohr energy, $\epsilon = -(E - E_B)$, of
the $1S$ level in kaonic hydrogen\cite{davies79}-\cite{bird83}.
Although these measurements should be a good test of the $\bar{K}N$
interaction slightly below threshold, the uncertainties in the
measurements make conclusions difficult (we see the statistical
uncertainty in Fig.~\ref{fig.H}, but not the apparent systematic
uncertainties).  Since at present there is distrust of a theory if it
agrees with these experiments, we eagerly await the new experiment in
progress with its promise of high precision and high
accuracy\cite{gill}.

In our calculation of the kaonic hydrogen state, we identify the
complex bound state energy as the $T$ matrix pole energy for the
combined Coulomb--plus--nuclear force problem\cite{landau83.1}, and
solve (\ref{pole}) for them.  Our predictions are shown in
Fig.~\ref{fig.H}. For all the CBM fits the calculated width $\Gamma$
is acceptable or slightly large, yet the shift $\epsilon$ is of
opposite sign to the data (all experimental shifts are to the more
bound). If the data are correct, then this implies that the the real
part of the $K^{-}p$ scattering amplitude (a sum of $I=0$ and $I=1$
amplitudes) is positive at $K^{-}p$ threshold even though the $I=0$
piece dips below zero at threshold.  The potential model of Schnick
and Landau\cite{schnick87} or its update by Tanaka and
Suzuki\cite{tanaka} have this property, that is, they agree in sign
with the data.  Although the sign changes in some of the cloudy bag
model's $\mbox{Re} f_{S01}$ are similar to those of the potential
models, none of our CBM fits have the proper combination of $I=0$ and
$I=1$ strengths to keep $\mbox{Re} f(K^{-}p) > 0$.  Although we have
not done it, we suspect that if this condition were made a requirement
of the search, such a solution would be found (presumably at the
expense of other data).

\section{Summary and Conclusions}

We have extended the $SU(3)$ cloudy bag model for the coupled
($\bar{K}N$, $\Sigma\pi$, $\Lambda\pi$) system from $S$ to $D$-waves
and thus to much higher energies.  The model has elementary quarks
inside a bag as well as an $SU(3)$ meson field inside and outside the
bag. While not derived from quarks, the meson field restore chiral
symmetry.  The model Hamiltonian contains a contact interaction which
generates direct meson-baryon scattering as well as an s--channel
interaction used to include elementary $\Lambda_{S01}(1405)$,
$\Sigma_{P13}(1520)$, and $\Lambda_{D03}(1385)$ resonances.  We
derived effective, energy--dependent separable potentials for use in
the Lippmann--Schwinger equation as well as some new numerical
approaches to solve the coupled Lippmann-Schwinger equation. The
parameters of the model were determined after extensive fits to
various scattering, reaction, branching ratio, and mass spectra data.

Our fitting and subsequent analyses indicate that the $\Sigma(1395)$
and $\Lambda(1520)$ are well described as elementary resonances.
Although their pole positions are renormalized by the contact
interaction and higher order scatterings from the Lippmann--Schwinger
equation to $31$ MeV below their bare masses, the resonance energies
(as determined by the complex energy poles of the $T$ matrix) are
within $2$ and $4$ MeV respectively of the tabulated values.

The cloudy bag model's description of the $\Lambda(1405)$ $S$-wave
resonance is less simple. While we agree with the previous conclusion
that the state is not an elementary, three--quark s-channel
resonance\cite{veit85.1,fink,dalitz91,alberg76,umino89}, we have also
found that it is less simple than the quasi-bound $\bar{K}N$, single
pole state produced by potential models\cite{schnick87,alberg76}. In
particular, there are two poles present in this channel, with the
resonant behavior near threshold arising from a pole above threshold
interfering with the threshold cusp.

Although the two sign changes in some of the cloudy bag model's
$\bar{K}N$ scattering amplitude near threshold are quite similar to
those of potential models which agree in sign with the strong
interaction shift in kaonic hydrogen, none of our CBM fits have quite
the proper combination of $I=0$ and $I=1$ strengths to keep $\mbox{Re}
f(K^{-}p) > 0$.  Agreement with the shift could be required as part of
the fitting procedure, but we suspect that this is best left for a
time when more acceptable data are available\cite{gill}.

In a general sense we conclude that the CBM with $S$, $P$, and $D$
waves is able to reproduce the $K^{-}p$ scattering and reaction cross
sections from $70 \rightarrow 513$ MeV/c ($1435 \rightarrow 1567$ MeV)
and either the $K^-p\rightarrow (\Sigma\pi\pi\pi, \Lambda\pi\pi\pi$)
mass spectra {\em or} the branching ratios
\begin{equation}
\gamma =  \frac{K^-p\rightarrow\Sigma^-\pi^+}
{K^-p\rightarrow\Sigma^+\pi^-}, \ \
R_c  =  \frac{K^-p\rightarrow\mbox{charged}}
{K^-p\rightarrow\mbox{all}}, \ \
R_n  =  \frac{K^-p\rightarrow\pi^0\Lambda}
{K^-p\rightarrow\mbox{neutral}}
\end{equation}
Fitting all three together appear too much to ask from such a simple
model.

We note that the fitted parameters appear reasonable. The
average bag radius $\overline{R} \simeq 1.22
\pm 0.14$ fm is about $0.1$ fm larger than the value found by
Veit et al.\cite{veit85.1}, but within the range $ 1.5 \geq R \geq
1.0$ fm given by Guidry\cite{guidry} (the range is for simple to
refined models, and our model with massless quarks is simple). If we
restrict our fit to low energy scattering and mass spectra data we
obtain $R=0.95$fm, which is not big and indicates that larger $R$
values arise from the effort to fit the high energy data.  The fitted
meson decay constants $(f^{I=0}, f_{K}^{I=1}, f_{\pi}^{I=0})
\simeq (99, 88, 91)$ MeV appear quite close to accepted\cite{guidry}
values $(f_{K}, f_{\pi}) = (112, 93)$ MeV, especially since $SU(3)$
predicts $f_{\pi} = f_{K}$.

In conclusion, we believe the cloudy bag model has success in
reproducing much data --- but clearly with limits.  The difficulties
may arise from trying to reproduce complicated energy dependences with
too simple a model. We have kept only linear terms in the meson field
and have assumed a rigid and spherical bag of one size for all
baryons. The bag does not recoil and the point--like mesons do not
arise from quarks. We expect this to cause difficulties at the higher
energies and higher momentum transfers. We have used the MIT bag
wavefunctions for massless quarks in a square well, and so have form
factors proportional to spherical Bessel functions. This too is rather
restrictive. We have looked at zero kinetic energy branching ratios,
but have included isospin--breaking effects only at the
Lippmann--Schwinger equation level not at the Hamiltonian level.
Clearly, all these effects may be important in a more sophisticated
study of branching ratios and of the kaonic hydrogen level shifts.
Further improvements appear worthwhile as do applications of the model
within a nuclear environment.

\acknowledgements

We wish to thank Al Stetz, Byron Jennings, Jeffrey Schnick, Tony
Thomas, Iraj Afnan, Gerry Miller, Mary Alberg, Larry Wilets, and
Ernest Henley for stimulating and illuminating discussions.  We also
wish to gratefully acknowledge support from the U.S. Department of
Energy under Grant \ DE-FG06-86ER40283, and the people at the National
Institute for Nuclear Theory, Seattle for their hospitality during
part of this work.

\appendix
\section{Reduction of $H_{s}$} \label{appa}

We convert $\hat H_s$  (\ref{eq:h.s.1}) into a more calculable form by
partial integration:
\begin{equation}
\hat H_s  =  -\int\!d^3x\left\{
\partial_\mu\left[\frac{\theta_v}{2f}\bar{q}\gamma^\mu\gamma_5
\vec\lambda q\cdot\vec\phi\right]
-\frac{(\partial_\mu\theta_v)}{2f}\bar{q}\gamma^\mu\gamma_5
\vec\lambda q\cdot\vec\phi\right.
\left. -\frac{\theta_v}{2f}\partial_\mu
\left[\bar{q}\gamma^\mu\gamma_5\vec\lambda
q\right]\cdot\vec\phi\right\} \label{hs}
\end{equation}
Application of the Dirac equation for the bag~(\ref{eq:dirac}) shows
the last term to vanish.  The space derivative in the first term is
converted to a surface integration over a infinitely large surface,
which in turn vanishes since the quark field is confined within the
bag.  The linear boundary condition~(\ref{eq:bc}) and the surface
delta function imply
$-(\partial_\mu\theta_v) \bar{q}\gamma^\mu\gamma_5
\vec\lambda q\cdot\vec\phi /2f =
-\frac{i}{2f}\delta_s\bar{q}\gamma_5\vec\lambda q\cdot\phi $ and this
yields:
\begin{equation}
\label{eq:h.s.2}
\hat H_s = \int\!d^3x\left[
\frac{i}{2f} \, \bar{q}\gamma_5\vec\lambda\cdot q\vec\phi\delta_s
-\frac{\theta_v}{2f} \, \partial_0
(\bar{q}\gamma^0\gamma_5\vec\lambda\cdot q\vec\phi)\right]
\end{equation}
We use this $H_{s}$ in the derivation of the potential to follow where
we separate off a vertex function $V$ in the Foch--space Hamiltonian
for $BM\leftrightarrow B_{0}'$:
\begin{eqnarray}
V_{0i}({\bf k})&=&
B_{0}'^{\dagger} \,\langle B_{0}'|\hat{H}_s| \alpha\rangle  \, B_0 \\
H_s &=& \sum_i\int\!d^3k\,[V_{0i}({\bf k})a_i({\bf k})
+V_{0i}^{\dagger}({\bf k})a_i^\dagger({\bf k})] \label{eq:H_s.s}
\end{eqnarray}

\section{$BM \leftrightarrow  \Lambda_{S01} (1405)$ Potential} \label{appb}

The quark model configuration for the bare $\Lambda_{S01}(1405)$ is
one u, d, and s quark in an SU(3) flavor singlet, with one $1p_{1/2}$
quark and two $1s$ quarks. The transition $BM\rightarrow \Lambda_{S01}
$ thus has one quark absorbing a meson and being excited from the $1s$
to $1p_{1/2}$ state. The quark wave functions are:
\begin{eqnarray}
\label{eq:q1s}
q_{1s}^M({\bf r}, t) &=& \frac{N_s}{\sqrt{4\pi}}
\left( \protect{
\begin{array}{c}
j_0(\omega_sr) \\
i \vec\sigma\cdot\hat r \, j_1(\omega_sr)
\end{array} }
\right)\chi_{\frac{1}{2}}^Me^{-i\omega_st} \,\theta(R-r) \\*[1ex]
q_{1p_{\frac{1}{2}}}^M({\bf r}, t) &=& \frac{N_{p1}}{\sqrt{4\pi}}
\left( \protect{
\begin{array}{c}
-\vec\sigma\cdot\hat r \, j_1(\omega_{p1}r)  \\
ij_0(\omega_{p1}r)
\end{array} }
\right) \chi_{\frac{1}{2}}^Me^{-i\omega_{p1}t} \,\theta(R-r)\\*[1ex]
N^{2}_{s,p1} &=& \frac{1} {2j_{0}^{2}(\omega_{s,p}R)R^{3}}
	\frac{\omega_{s,p} R} {\omega_{s},p \mp1}
\end{eqnarray}
Here $\chi$ is the spin-flavor wave function of the quark, $j_0$ and
$j_1$ are spherical Bessel functions, $(\omega_s, \omega_{p1})
\approx (2.04/R, 3.81/R)$ are the energies of $1s$ and $1p_{1/2}$
states, and the $N$'s are normalization constants\cite{hey77}.  The
$BM \leftrightarrow \Lambda_{S01} (1405)$ vertex function needed in
the Foch--space Hamiltonian $H_{s}$ (\ref{eq:H_s.s}) and the form
factor for the Hamiltonian are accordingly:
\begin{eqnarray}
V_{0i}^{(S01)}({\bf k})&=&
\Lambda_{1/2}^{\dagger} \, v_{ \Lambda_{S01} \alpha}({\bf k})\, B_0 \\
v_{ \Lambda_{S01} \alpha}({\bf k}) & = &
	\langle \Lambda_{S01} |H_s| \alpha\rangle  \\
&=&  \frac{-1}{2f}\frac{ {}^{sf}\!\langle \Lambda_{S01}
|\lambda_i| B\rangle ^{sf} }
{ \sqrt{(2\pi)^32\omega_{k}}}
N_sN_{p1} \left\{
2R^2j_0(\omega_sR)j_0(\omega_{p1}R)j_0(kR)
 \right. \nonumber \\ & & + (\omega_{k}+\omega_s-\omega_{p1})
\left. \int_0^R\!dr\,r^2\left[j_0(\omega_sr)j_0(\omega_{p1}r)
+j_1(\omega_sr)j_1(\omega_{p1}r)\right]j_0(kr) \right\}
\label{eq:H_s1.vertex}
\end{eqnarray}
The spin-flavor matrix element
${}^{sf}\!\langle \Lambda_{S01} |\lambda_i
| B\rangle ^{sf}$ is evaluated with the
Wigner-Eckart theorem:
\begin{equation}
^{sf}\!\langle \Lambda_{S01} |\lambda_i| B\rangle ^{sf} =
\lambda_\alpha^\Lambda \langle {I_B} {i_B}; {I_M} {i_M}
| {I_B}  {I_M} ; 00 \rangle
\label{eq:lambda}
\end{equation}
where the bracket on the right-hand side is a Clebsch-Gordan
coefficient. The coupling constants $\lambda_\alpha^\Lambda$ are
calculated using the SU(6) quark wave function\cite{veit85.1,gibson}
and are listed in Table~\ref{tab.couple}.  The potential
$v^{(S01)}({\bf k'},\mu'; {\bf k},\mu)$ corresponds to meson--baryon
scattering through an intermediate $ \Lambda_{S01} $ state,
Fig.~\ref{fig.res}:
\begin{eqnarray}
v_{\beta\alpha}^{(S01)} &=&
\langle \beta |H_s |  \Lambda_{S01}
\rangle \frac{1}{E-M_{S01}}\langle \Lambda_{S01} |H_s| \alpha\rangle  \\
 & = &
\frac{\delta_{\mu\mu'}\lambda^\Lambda_\beta\lambda^\Lambda_\alpha
 \langle {I_{B'}} {i_{B'}}; {I_{M'}} {i_{M'}}
| I_{B'}  {I_{M'}}; {0} {0} \rangle }{E-M_{S01}}
\frac{ \langle {I_B} {i_B} ; {I_M} {i_M}
| {I_B} {I_M}; {0}{0} \rangle}{64f^{2}\pi^{3}}
\frac{u_{ \Lambda_{S01} \beta}(k)u_{ \Lambda_{S01} \alpha}(k)}
{\sqrt{ \omega_{k'}\omega_{k}}} \label{potS01}
\end{eqnarray}
The partial--wave matrix projection of this potential via
(\ref{eq:vproject}) yields (\ref{eq:v.p13}).

\section{$BM \leftrightarrow \Lambda_{D03} (1520)$ Potential} \label{appc}

The $\Lambda_{D03}(1520)$ has one u, d, and s quark in an $SU(3)$
singlet with total spin $3/2$ and wavefunction $1s^2\,1p_{3/2}$. The
$\bar{K}N \rightarrow \Lambda_{D03} $ transition thus has one quark
absorbing the antikaon and changing from $1s$ to the $1p_{3/2}$
(flavor may also change). The vertex function for the Foch--space
Hamiltonian (\ref{eq:H_s.s}) and the $1p_{3/2}$ quark wave function
are:
\begin{eqnarray}
V_{0i}^{(D03)}({\bf k}) &=&
\Lambda_{3/2}^{\dagger}\int\!d^3x e^{i{\bf k\cdot r}}\,
\frac{{}^{sf}\!\langle\Lambda_{D03} |
\delta_s\bar{q}_{p}\gamma_5\lambda_iq_{s}
+(\omega_s+\omega_{k}-\omega_{p3})
\bar{q}_{p}\gamma^0\gamma_5\lambda_iq_{s}| B\rangle ^{sf} }
{-2fi\sqrt{(2\pi)^32\omega_{k}}}
B_0 \label{eq:H_s.d} \\*[1ex]
\label{eq:q1p3}
q_{1p_{3/2}}^M({\bf r}, t) &=& N_{p3}
\left( \protect{
\begin{array}{c}
j_1(\omega_{p3}r) \\
ij_2(\omega_{p3}r)( \mbox{\boldmath$\sigma$}\cdot\hat r)
\end{array} }
\right)\theta(R-r)e^{-i\omega_{p3}t}
{\cal Y}_{1\frac{3}{2}}^M(\theta, \phi)
\end{eqnarray}
Here ${\cal Y}^{M}_{lj}$ is the spin-angle function, $N_{p3}$ is the
$1p_{3/2}$ normalization constant\cite{hey77}, and $\omega_{p3}\approx
3.20/R$ is the energy level of the $1p_{3/2}$ state.  We express the
spin--angle function in terms of spherical harmonics and Pauli
spinors, substitute $| \Lambda_{D03} \rangle ^{sf} =
\sum_m  \langle {1} {m}; {\frac{1}{2}} {M-m}
| {1} {\frac{1}{2}}; {\frac{3}{2}} {M} \rangle Y_{1m}| \Lambda_{S01}
\rangle ^{sf} $ for the $\Lambda_{D03} $ spin-flavor wavefunction (the
same as $ \Lambda_{S01} $) to obtain:
\begin{equation}
{}^{sf} \langle \Lambda_{S01} (\mu-m)
|\lambda_i\sigma_q| B(\nu)\rangle ^{sf}
 =  -\sqrt{3}\,\lambda_\alpha^\Lambda
 \langle {\frac{1}{2}} {\nu}; {1} {q}
| {\frac{1}{2}} {1}; {\frac{1}{2}} {\mu-m} \rangle
 \langle {I_B} {i_B}; {I_M} {i_M} | {I_B} {I_M}; {0} {0} \rangle
\end{equation}
\begin{eqnarray}
\vspace{-4ex}
v_{\Lambda_{D03} \alpha} (\mu, \nu) &=&
-\frac{1}{2f}\frac{1}{\sqrt{3}}
\frac{4\pi N_sN_{p3}}{\sqrt{(2\pi)^32\omega_{k}}}\sum_{mq}\sum_{LM}i^L(-1)^q
 \langle {1} {m}; {\frac{1}{2}} {\mu-m} | {1}
{\frac{1}{2}};{\frac{3}{2}} {\mu} \rangle Y_{LM}^*(\hat k) \nonumber
\\ & &
\times\,{}\,^{sf}\!\langle \Lambda_{S01} (\mu-m)
|\lambda_i\sigma_q| B(\nu)\rangle ^{sf}
\int\!d^3xY_{1-q}(\hat r)Y^*_{1m}(\hat r)Y_{LM}(\hat r)j_L(kr)
\nonumber \\
 & & \times\, \{\delta_s[j_1(\omega_sr)j_1(\omega_{p3}r)
+j_0(\omega_sr)j_2(\omega_{p3}r)] \nonumber \\ & & \mbox{\hspace{1em}}
+\theta_v(\omega_s+\omega_{k}-\omega_{p3})
[j_1(\omega_sr)j_1(\omega_{p3}r)-j_0(\omega_sr)j_2(\omega_{p3}r)]\} \\
& = & \frac{1}{4\pi f \sqrt{3\omega_{k}} }
\sum_{LM}\sum_{mq}(-1)^mi^L\sqrt{2L+1}
Y_{LM}^*(\hat k)u_{\Lambda_{D03} \alpha}^{(L)}(k)
\langle {L} {0}; {1} {0} | {L}  {1};{1} {0} \rangle \nonumber \\
 & & \times\,{}^{sf}\!\langle \Lambda_{S01} (\mu-m)
|\lambda_i\sigma_q| B(\nu)\rangle ^{sf}
     \langle {1} {m}; {\frac{1}{2}} {\mu-m}
| {1} {\frac{1}{2}};{\frac{3}{2}} {\mu} \rangle
\langle {L} {M}; {1} {-m} | {L} {1};{1} {q} \rangle\label{eq:u3} \\
u_{\Lambda_{D03} \alpha}^{(L)}(k) &=& -N_sN_{p3} \left[
 2R^2j_0(\omega_sR)j_1(\omega_{p3}R)j_L(kR) \right. \nonumber\\
 & & + \left.(\omega_s+\omega_{k}-\omega_{p3}) \int_0^R\!dr\,r^2j_L(kr)
[j_1(\omega_sr)j_1(\omega_{p3}r)-j_0(\omega_sr)j_2(\omega_{p3}r)]
\right]
\end{eqnarray}
The coupling constants $\lambda^\Lambda_\alpha$ are the same as in the
$BM\leftrightarrow \Lambda_{S01} $ case (Table~\ref{tab.couple}), and after
lengthy algebraic manipulation of the Clebsch-Gordan coefficients and
$6j$ symbols, we obtain the vertex function:
\begin{eqnarray}
v_{\Lambda_{D03} \alpha}({\bf k};\mu, \nu)& = &
\frac{\lambda^\Lambda_\alpha}{4\pi f\sqrt{\omega_{k}}}
 \langle {I_B} {i_B}; {I_M} {i_M} | {I_B} {I_M};{0} {0}  \rangle
\nonumber \\
& & \times\,\sum_{LM}\sum_{mq}(-1)^m i^L\sqrt{2L+1}
Y^*_{LM}(\hat k)
\langle {L} {0}; {1} {0} | {L} {1} ;{1} {0} \rangle
 u_{\Lambda_{D03} \alpha}^{L}(k)  \nonumber \\
& & \times\,
 \langle {1} {m}; {\frac{1}{2}} {\mu-m}
| {1} {\frac{1}{2}}; {\frac{3}{2}} {\mu} \rangle
 \langle {L} {M}; {1} {-m} | {L} {1}; {1} {q} \rangle \,
 \langle {\frac{1}{2}} {\nu} ; {1} {q}
| {\frac{1}{2}} {1}; {\frac{1}{2}} {\mu-m} \rangle
\nonumber\\
& = & \frac{\lambda^\Lambda_\alpha u_{\Lambda_{D03} \alpha}(k)}
{4\pi f\sqrt{\omega_{k}}}
 \langle {I_B} {i_B}; {I_M} {i_M} | {I_B} {I_M}; {0} {0} \rangle
 \langle {\frac{1}{2}} {\nu}; {2} {\mu-\nu}
| {\frac{1}{2}} {2}; {\frac{3}{2}} {\mu}  \rangle
Y_{2(\mu-\nu)}^*(\hat k)
\end{eqnarray}
The separable potential $v^{(D03)}$ corresponds to meson--baryon
scattering through the intermediate $\Lambda_{D03} $,
Fig.~\ref{fig.res}, i.e. using $\Lambda_{D03} $ for $B_{0}$ in
(\ref{pot}):
\begin{eqnarray}
v_{\beta\alpha}^{(D03)}({\bf k'}, {\bf k})
& = &\sum_\nu\langle \beta(\mu')|H_s
| \Lambda_{D03} (\nu)\rangle  \frac{1}{E-M_{D03}}
\langle \Lambda_{D03} (\nu)|H_s  | \alpha(\mu)\rangle  \nonumber \\
& = & \frac{\lambda_\beta^\Lambda\lambda_\alpha^\Lambda
u_{\Lambda_{D03} \beta}(k')u_{\Lambda_{D03} \alpha}(k) }
{16\pi^2f^{2}\sqrt{\omega_{k}\omega_{k'}} }
\frac{  \langle {I_{B'}} {i_{B'}}; {I_{M'}} {i_{M'}}
| {I_{B'}} {I_{M'}};{0}{0} \rangle \, \langle {I_B} {i_B}; {I_M} {i_M}
| {I_B} {I_M}; {0}{0}\rangle } {E-M_{D03}}\nonumber \\ & & \times\,
\sum_\nu Y_{2(\nu-\mu)}^*(\hat k)Y_{2(\nu-\mu')}(\hat k')
  \langle {\frac{1}{2}} {\mu}; {2} {\nu-mu}
| {\frac{1}{2}} {2}; {\frac{3}{2}} {\nu} \rangle
 \langle {\frac{1}{2}} {\mu'}; {2} {\nu-mu'}
| {\frac{1}{2}} 2; {\frac{3}{2}} {\nu}  \rangle
\label{eq:v.d}
\end{eqnarray}
The partial wave projection  of $v^{(D03)}$ follows
pages of algebra and yields (\ref{eq:v.d03}).

\section{$BM \leftrightarrow \Sigma_{P13}(1385)$ Potential} \label{appd}

The bare $\Sigma_{P13}(1385)$ is a single u, d, and s quarks in a
$\underline{10}$ representation of SU(3), with all quarks in the $1s$
state. The transition $BM\rightarrow\Sigma_{P13}$ thus has one quark
changing its flavor--spin state after absorbing a meson.  The vertex
function and form factor for the Foch--space Hamiltonian
(\ref{eq:H_s.s}) are accordingly:
\begin{eqnarray}
V_{0i}^{(p)}({\bf k}) &=&
\Sigma_{3/2}^{0*\dagger} \, v_{\Sigma_{P13}\alpha}({\bf k}) \, B_0\\
v_{\Sigma_{P13}\alpha}({\bf k}) &=&
\Sigma_{3/2}^{0*\dagger} \, \langle\Sigma_{P13}|H_s| \alpha\rangle  B_0
\label{eq:vs.p} \\
&=&
\frac{1}{2f}\int\!\frac{d^3r}{\sqrt{(2\pi)^32\omega_{k}}}e^{i{\bf k\cdot r}}
\langle\Sigma_{P13}|i\delta_s\bar{q}_{1s}\gamma_5\lambda_iq_{1s}
+i\theta_v\omega_{k}\bar{q}_{1s}\gamma^0\gamma_5\lambda_iq_{1s}| B
\rangle
\label{eq:vs.p.1}\\
&=& \label{52}
\frac{1}{2f}\frac{N_s^2}{4\pi}
(-2)\int\!\frac{d^3r}{\sqrt{(2\pi)^32\omega_{k}}} e^{i{\bf k\cdot
r}}j_0(\omega_sr)j_1(\omega_sr)\delta_s
{}^{sf}\!\langle\Sigma_{P13}|\lambda_i \mbox{\boldmath$\sigma$}\cdot\hat r| B
\rangle ^{sf} \\ & = &
\frac{1}{2f}\frac{N_s^2}{4\pi}\frac{-2}{\sqrt{(2\pi)^32\omega_{k}}}
\int\!d^3r\,\delta_s4\pi
\sum_{LM}i^Lj_L(kr)Y_{LM}^*(\hat k)Y_{LM}(\hat r) \nonumber \\
 & & \times\, j_0(\omega_sr)j_1(\omega_sr)
\sum_q\sqrt{\frac{4\pi}{3}}Y_{1q}^*(\hat r)
{}^{sf}\!\langle\Sigma_{P13}
|\lambda_i\sigma_q| B\rangle ^{sf} \nonumber \\
&=& \frac{-iN_s^2R^2}{4f\sqrt{\pi^3 \omega_{k}}}
\sum_qY_{1q}^*(\hat k)j_0(\omega_sR)j_1(\omega_sR)j_1(kR)
{}^{sf}\!\langle\Sigma_{P13}|\lambda_i\sigma_q| B\rangle ^{sf}
\end{eqnarray}
To obtain the vertex function, we substitute (\ref{eq:q1s}) for the
$1s$ quark wave function, substitute the partial wave expansion of the
plane wave, substitute the explicit form for
$ \mbox{\boldmath$\sigma$}\cdot\hat r$, and use the Wigner-Eckart theorem:
\begin{eqnarray}
 \mbox{\boldmath$\sigma$}\cdot\hat r &=&
\sqrt{\frac{4\pi}{3}}\sum_qY_{1q}^*(\hat r)\sigma_q \, , \ \ \
\sigma_{\pm 1,0} =  \mp\frac{1}{\sqrt{2}}(\sigma_x\pm\sigma_y), \
\sigma_z \label{eq:sgmpm} \\
 {}^{sf}\!\langle\Sigma_{P13}(\mu)|\lambda_i\sigma_q| B(\nu)\rangle ^{sf}
&=& \lambda_{\alpha}^\Sigma \langle {\frac{1}{2}} {\nu}; {1} {q}
| ; {\frac{1}{2}} {1}{\frac{3}{2}} {\mu} \rangle
 \langle  {I_B} {i_B}; {I_M} {i_M}| {I_B} {I_M}; {1}{0} \rangle\\
v_{\Sigma_{P13}\alpha}({\bf k}; \mu, \nu) &=&
\frac{-i\lambda_\alpha^\Sigma N_s^2R^2 }{2\pi f \sqrt{3\omega_{k}} }
 \langle {\frac{1}{2}} {\nu}; {1} {q}
|  {\frac{1}{2}} {1}; {\frac{3}{2}}{\mu} \rangle
 \langle  {I_B} {i_B}; {I_M} {i_M} | {I_B} {I_M}; {1}{0}\rangle \nonumber \\
& & \times \, Y_{1(\mu-\nu)}^*(\hat k)
j_0(\omega_sR)j_1(\omega_sR)j_1(kR)
\label{eq:vsp}
\end{eqnarray}
The calculated coupling constants $\lambda_\alpha^\Sigma$ are given in
Table~\ref{tab.couple}.  The potential $v^{(P13)}({\bf k'},\mu'; {\bf
k},\mu)$ is similar to $v^{(D03)}$, but now with an intermediate
$\Sigma_{P13}$ (Fig.~\ref{fig.res}):
\begin{eqnarray}
v_{\beta\alpha}^{(P13)}({\bf k'},{\bf k}) &=&
\sum_\nu\langle\beta(\mu')|H_s
| \Sigma_{P13}(\nu)\rangle  \frac{1}{E-M_{P13}}
\langle\Sigma_{P13}(\nu)|H_s| \alpha(\mu)\rangle  \\
& = &
\frac{1} {12 \pi^2 f^{2} \sqrt{\omega_{k'}\omega_{k}}}
\frac{\left(N_s^2R^2 j_0(\omega_sR)j_1(\omega_sR)\right)^2}{E-M_{P13}}
j_1(k'R)j_1(kR) \lambda_\beta^\Sigma\lambda_\alpha^\Sigma \nonumber \\
& &\times\sum_\nu Y_{1(\nu-\mu)}^*(\hat k)Y_{1(\nu-\mu')}(\hat k')
 \langle {\frac{1}{2}} {\mu}; {1} {\nu-mu}
| {\frac{1}{2}} {1}; {\frac{3}{2}}{\nu}  \rangle
 \langle {\frac{1}{2}} {\mu'}; {1} {\nu-mu'}
| {\frac{1}{2}} {1}; {\frac{3}{2}}{\nu} \rangle\nonumber \\
& & \times \langle {I_{B'}} {i_{B'}}; {I_{M'}} {i_{M'}}
| {I_{B'}} {I_{M'}} ; {1}{0} \rangle
 \langle {I_B} {i_B}; {I_M} {i_M}| {I_B} {I_M}; {1}{0} \rangle
\end{eqnarray}
Its partial wave projection via (\ref{eq:vproject}) yields
(\ref{eq:v.p13}).

\section{Contact Potentials} \label{space}

The space--derivative part of the contact potential is:
\begin{eqnarray}
v_{\beta\alpha}^{(cs)}({\bf k'}, {\bf k}) &=&
\frac{i {}^{sf}\!\langle B'|\int\!d^3x\,\theta_v\bar{q}_{1s}
f_{i'ij}\lambda_j({\bf k'+k})\cdot \mbox{\boldmath$\gamma$}
e^{i{\bf(k-k')\cdot r}}q_{1s}| B\rangle ^{sf} } {8 f^2 (2\pi)^3
\sqrt{\omega_{k}\omega_{k'}}} \\
& = &
\frac{iN_s^2}{4\pi}\int_0^R\!drd\Omega_r\,rj_0(\omega_sr)j_1(\omega_sr)
\,{}^{sf}\langle B'
|f_{i'ij}\lambda_j   \mbox{\boldmath$\sigma$}| B\rangle ^{sf}
\nonumber \\
& & \hspace{3ex} \times \frac{ e^{-i{\bf k'\cdot r}}({\bf r\times k})e^{i{\bf
k\cdot r}} +({\bf r\times k'})e^{-i{\bf k'\cdot r}}e^{i{\bf k\cdot r}}
} { 4 (2\pi)^3 f^2 \sqrt{\omega_{k} \omega_{k'}} }
\end{eqnarray}
where we have substituted for the quark wave functions $q_{1s}$ and
$\bar{q_{1s}}$. Since ${\bf r\times k}$ and
${\bf r\times k'}$ are the angular momentum operators acting on
$e^{i{\bf k\cdot r}}$ and $e^{i{\bf k\cdot r}}$, substituting the
partial wave expansions of the plane waves yields:
\begin{eqnarray}
v_{\beta\alpha}^{(cs)}({\bf k'}, {\bf k}) & = &
\frac{iN_s^2 {\cal A} }
{f^2 (2\pi)^2 \sqrt{\omega_{k} \omega_{k'}}}
\sum_{lmm'}Y^*_{lm}(\hat k)Y_{lm'}(\hat k')
\int_0^R\!dr\,j_0(\omega_sr)j_1(\omega_sr)
j_l(kr)j_l(k'r) \label{eq:vcs.1} \\
{\cal A} &=& {}^{sf}\!\langle B'(\mu')|f_{i'ij}\lambda_j
 \mbox{\boldmath$\sigma$}  | B(\mu)\rangle ^{sf}\,\int\!d\Omega_r
Y^*_{lm'}{\bf L}Y_{lm}  \nonumber \\
& = & \sum_q(-1)^q\int\!d\Omega_rY^*_{lm'}L_qY_{lm}\,
{}^{sf}\!\langle B'(\mu')|f_{i'ij}\lambda_j\sigma_{-q}| B(\mu) \rangle ^{sf}
\nonumber \\
& = & {}^f\!\langle B'||f_{i'ij}\lambda_j  \mbox{\boldmath$\sigma$}
|| B^ \rangle f
\,\sum_j(-1)^{\frac{1}{2}+j+l}\sqrt{l(l+1)(2l+1)}
\left\{\begin{array}{ccc}
\frac{1}{2} & \frac{1}{2} & 1 \\
l & l & j \end{array}
\right\} \nonumber \\
& & \hspace{3ex} \times  \langle {\frac{1}{2}} {\mu}; {l} {m}
| {\frac{1}{2}} {l}; {j}{m'+\mu'} \rangle
 \langle {\frac{1}{2}} {\mu'}; {l} {m'}
|  {\frac{1}{2}} {l};{j}{m'+\mu'} \rangle
\label{eq:a6j}
\end{eqnarray}
where we have used the tensor notation for $\sigma_{-q}$ and
$L_{q}$ as in (\ref{eq:sgmpm}), the  Wigner-Eckart
theorem, and many manipulations.
The calculated coupling constants $\lambda_{\beta\alpha}^{s,I}$ are
given in Table~\ref{tab.couple}.  After substituting all these
relations back, we obtain:
\begin{eqnarray}
v_{\beta\alpha}^{(cs)}({\bf k'},{\bf k})
& = & - N_s^2
\sum_{I,jM,lmm'} \lambda_{\beta\alpha}^{s,I}
\langle {I_B} {i_B}; {I_M} {i_M} | {I_B}  {I_M};{I}{0} \rangle
 \langle {I_{B'}} {i_{B'}}; {I_{M'}} {i_{M'}}
| {I_{B'}} {I_{M'}}; {I}{0}\rangle  \nonumber \\
& & \times  \langle {\frac{1}{2}} {\mu}; {l} {m}
|  {\frac{1}{2}} {l} ; {j}{M}\rangle
 \langle {\frac{1}{2}} {\mu'}; {l} {m'}
| {\frac{1}{2}} {l}; {j}{M} \rangle
Y^*_{lm}(\hat k)Y_{lm'}(\hat k')
\sqrt{6l(l+1)(2l+1)}  \nonumber \\
& & \times (-1)^{j+l+\frac{1}{2}} \int_0^R
\frac{ dr\,r j_0(\omega_sr) j_1(\omega_sr) j_l(kr)j_l(k'r)
 } {2f^2
\pi^2\sqrt{\omega_{k}\omega_{k'}}} \left\{\begin{array}{ccc}
\frac{1}{2} & \frac{1}{2} & 1  \\
l & l & j \end{array}
\right\} \label{eq:vcs.f}
\end{eqnarray}
The equations for $v^{(cs)}$ and $v^{(ct)}$ have the same form as
equations (2.8) and (2.9) in Ref. \cite{veit85.2}, but the coupling
constants (spin-flavor matrix elements) in Table~\ref{tab.couple} are
different.  Finding the partial wave matrix elements of the potential
$v^{(cs)}$ is more complicated.  We start with the definition
(\ref{eq:tlj}),  apply the orthogonal relations of the spherical
harmonics, the properties of the Clebsch-Gordan coefficients, and
lengthy manipulations to obtain (\ref{eq:pw.cs}).

\section{Extended Haftel-Tabakin Technique}\label{app.num}

We give here our extension of the Haftel-Tabakin
technique\cite{haftel70} for solving the Lippmann-Schwinger
equation~(\ref{eq:ls1d}). For each channel $\gamma$ the integrand in
(\ref{eq:ls1d}) contains an integrable singularity at the ``on-shell''
channel momentum $k_{0\gamma} \equiv  k_{0}$ defined in
(\ref{channelE}). To permit numerical integration this singularity is
removed by subtracting a term from the integrand which leaves the
integrand nonsingular\cite{haftel70}, and then adding in the integral
of the subtracted term:
\begin{eqnarray}
\int_0^\infty\!dp\,
\frac{p^2V(k',p)T(p,k)}
{E-E(p)+i\epsilon}
 & = &
\int_0^\infty\!dp\,
\left[\frac{p^2V(k',p)T(p,k)}
{E-E(p)+i\epsilon}
-\frac{2\mu k_{0}^2V(k', k_{0})
T(k_{0}, k)}{k_{0}^2-p^2+i\epsilon}
\right]
\nonumber \\
& & + 2\mu k_{0}^2
V(k',k_{0})
T(k_{0},k)
\int_0^\infty\!dp\,\frac{1}{k_{0}^2-p^2+i\epsilon}
\label{eq:sub}\\
& = &
\int_0^\infty\!dp\,
\left[\frac{p^2V(k', p)T(p,k)}
{E-E(p)}
-\frac{\pi\mu k_{0}^2V(k',k_{0})
T(k_{0},k)}{k_{0}^2-p^2}
\right]
\nonumber \\
& & -\pi i \mu k_{0}
V(k',k_{0})
T(k_{0},k)
\label{eq:sub2}
\end{eqnarray}
where the RHS of Eq.~(\ref{eq:sub}) is evaluated analytically and
where $\mu \equiv \mu_\gamma $ is the relativistic ``reduced mass''
for channel $\gamma$:
\begin{equation}
\mu_\gamma = \frac{1}{2}\left.\frac{dp^2}{dE_\gamma}\right|_{E_\gamma
= E} = \frac{E_{1}(k_{0\gamma})
E_{1}(k_{0\gamma})}{E_{1}(k_{0\gamma})+E_{1}(k_{0\gamma})}
\label{eq:reduced}
\end{equation}
We solve the integral equations (\ref{eq:ls1d}) by approximating the
integrals as sums over $N$ Gaussian grid points $\{p_i | i = 1, N\}$
with weights $\{w_i | i = 1, N\}$:
\begin{eqnarray}
T_{\beta\alpha}(k', k) & = &
V_{\beta\alpha}(k', k) -\sum_\gamma 2i\mu_\gamma k_{0\gamma}
V_{\beta\gamma}(k', k_{0\gamma})
T_{\gamma\alpha}(k_{0\gamma}, k)
\nonumber \\
& & + \frac{2}{\pi}\sum_{\gamma,i}
\left[\frac{p_i^2V_{\beta\gamma}(k', p_i)T_{\gamma\alpha}(p_i, k)}
{E - E_\gamma(p_i)}
-\frac{2\mu_\gamma k_{0\gamma}^2V_{\beta\gamma}(k', k_{0\gamma})
T_{\gamma\alpha}(k_{0\gamma}, k)}{k_{0\gamma}^2-p_i^2}
\right]
\label{eq:tgrid}
\end{eqnarray}
We convert (\ref{eq:tgrid}) to the set of linear equations,
\begin{equation}
T_{n\beta,m\alpha} = V_{n\beta,m\alpha}
+ \sum_{\gamma=1}^{N_{c}}\sum_{i=1}^{N+1}
V_{n\beta,i\gamma}G_{i\gamma}T_{i\gamma, m\alpha} \label{eq:ls}
\end{equation}
by defining the supematrix elements for $G$, $V$, and $T$:
\begin{eqnarray}
V_{n\beta,m\alpha} &=&
\left\{ \protect{\begin{array}{clcl}
V_{\beta\alpha}(p_n, p_m) & \mbox{\ \ \ ($n = 1\, N$}, &
& \mbox{$m = 1\, N$)} \\
V_{\beta\alpha}(k_{0\beta}, p_m) & \mbox{\ \ \ ($n = N+1$}, &
& \mbox{$m = 1\, N$)} \\
V_{\beta\alpha}(p_n, k_{0\alpha}) & \mbox{\ \ \ ($n = 1\, N$}, &
& \mbox{$m = N+1$)} \\
V_{\beta\alpha}(k_{0\beta}, k_{0\alpha}) & \mbox{\ \ \ ($n = N+1$}, &
& \mbox{$m = N+1$)}
\label{eq:v.def}
\end{array}} \right. \\
G_{n\gamma} &=& \left\{\protect{
\begin{array}{ll}
2 w_n p_n^2/(\pi(E-E_\gamma(p_n)) &
\mbox{\ \ \ ($n = 1\, N$)} \\
-2\mu_{\gamma}\sum_{i=1}^N \left(2 w_ik_{0\gamma}^2/
(\pi (k_{0\gamma}^2-p_i^2)\right)
-ik_{0\gamma} 2\mu_{\gamma} &
\mbox{\ \ \ ($n = N+1$)}
\end{array}}\right.
\label{eq:g.def}
\end{eqnarray}
Here $T_{n\beta,m\alpha}$ are the $N_{c}^2(N+1)^2$ unknowns for
$N_{c}$ channels and $N$ Gaussian grid points (the $+1$ arising from
the on--shell point).  The potential $V$ is taken as the sum of the
$5$ terms in (\ref{pot}), $N_{c}=5$ for the five independent isospin
channels in (\ref{eq:ibasis}), and we use $24$ or $32$ Gaussian grid
points in each channel.  We then solve the $N_{c}^{2}(N+1)^{2}$
coupled equations (\ref{eq:ls}) by rearranging the matrix equation:
\begin{eqnarray}\label{eq:t.m2}
[T] &=& [V] + [VG][T]\\
\hspace{3ex} \Rightarrow & &[1-VG][T] = [V\label{eq:t.m}\\
\hspace{5ex} \Rightarrow & & [T] = [1-VG]^{-1}[V]\label{eq:t.m3}
\end{eqnarray}

Because the solution for $T$ is numerically intensive, we have made a
number of improvements to our former technique. First, while we
previously determined the actual  inverse $[1-VG]^{-1}$ and evaluated
(\ref{eq:t.m3}), we now solve (\ref{eq:t.m}) for $T$ using Gaussian
elimination.  This is $\sim 30$\% faster than matrix inversion. Next
we obtain an even greater savings by utilizing the symmetry of the
potential $V$ and the diagonal nature of the Green's function
$G$ to reformulate (\ref{eq:t.m}) as
\begin{equation}
[G^{-1} - V] [G T] = [V]
\end{equation}
Since $[G^{-1} - V]$ is symmetric, we solve a symmetric linear system
of equations for $[GT]$ --- which is much faster than a general
system. From this $[GT]$ we easily obtain the value of $[T]$ since
$[G]$ is a diagonal and its inverse is quick to compute.

If there are some elements of $G$ which vanish, the reformulation is a
little more complicated. For example, we assume the last element of
$G$ vanishes while none of the others do (if it is not the last
element, we can always rearrange columns and rows to make it last).
The linear system is now:
\begin{equation}
\label{eq:subvg1}
\left[ \begin{array}{cc}
    1 - vg & 0 \\
    -v_rg  & 1
\end{array}\right]
\left[ \begin{array}{cc}
    t      & t_c \\
    t_r    & t_M
\end{array}\right]
=
\left[\begin{array}{cc}
    v  & v_c \\
    v_r & v_M
\end{array}\right]
\end{equation}
Here $v$ is an $(M-1)\times(M-1)$ {\em symmetric} matrix, $g$ is a
$(M-1)\times(M-1)$ diagonal matrix, $t$ is a $(M-1)\times(M-1)$
matrix, $v_{r}$ and $t_{r}$ are $(M-1)\times 1$ row matrices, $v_c$
and $t_c$ are $(M-1)\times 1$ column matrices, and $v_{M}$ and $t_{M}$
are the lower right--hand corner elements of $V$ and $T$. We split
(\ref{eq:subvg1}) into the equations:
\begin{eqnarray}
(1-vg)t        & = & v   \\
(1-vg)t_c      & = & v_c \\
-v_rgt + t_r   & = & v_r \\
-v_rgt_c + t_M & = & v_M
\end{eqnarray}
and solve for $t$ and $t_c$ with the {\em symmetric} matrix $v$ (this
is fast). We then solve for $t_r$ and $t_M$ if needed.  If there are
several elements in $G$ which vanish, we follow the same procedure
recursively until we are left with a symmetric system to solve. In
practice, the direct LU decomposition method took 1214 second on an
IBM RS/6000 Model-530, while the method using symmetry took 493
seconds.



\begin{figure}
\caption{Feynman diagrams for  meson interactions in the cloudy bag
model, (A): via the s--channel Hamiltonian $H_{s}$, (B): via the
contact Hamiltonian $H_{c}$.  Dashed lines represent mesons and solid lines
fermions (quarks or baryons).}
\label{fig.feyn} 
\end{figure}

\begin{figure}
\caption{The Feynman diagram used to generate potentials corresponding
to intermediate resonance excitation via the Hamiltonian $H_{s}$.}
\label{fig.res} 
\end{figure}

\begin{figure}
\caption{The $K^{-}p$ scattering cross sections, reaction cross
sections, and $\Sigma\pi$ mass spectrum calculated with the $S$ wave
fits 1, 2, and Triumf (A).  As is evident, fit 1 was not adjusted to
fit the mass spectrum.  The cusps right below $100$ MeV/c arise from the
opening of the $\bar{K}^{0}N$ channel, and the peaks near $400$ MeV/c
arise from the $\Lambda(1520)$ $D$--wave resonance (not included in
these $S$-wave fits).  The cross section data are $\circ$: Cibrowski
et al.\protect\cite{ciborowski82}, $\Box$: Kim\protect\cite{kim},
$\diamond$: Mast et al.\protect\cite{mast}, $\bigtriangleup$: Sakitt
et al.\protect\cite{sakitt65}, $\triangleleft$: Watson et
al.\protect\cite{watson63}, $\bigtriangledown$: Evans et
al.\protect\cite{evans83}, $\triangleright$: Kittel et
al.\protect\cite{kittel66}, and $\times$: Bangerter et
al.\protect\cite{bangerter81}. The $\Sigma\pi$ mass spectrum is from
Hemingway\protect\cite{hemingway85}.}
\label{fig.fit1+2}
\end{figure}

\begin{figure}
\caption{The $K^{-}p$ scattering cross sections,
reaction cross sections, $\Sigma\pi$ mass spectrum, and
$\pi^{0}\Lambda^{0}$ mass spectrum calculated with parameters of fits
3 and 4.  Both fits include $S$, $P$, and $D$ waves, but fit 4 was not
adjusted to branching ratios (shown in Fig.~\protect\ref{fig.br}). The
cross section data are the same as in Fig.~\protect\ref{fig.fit1+2}
with the addition of the $\pi^{0}\Lambda^{0}$ mass spectrum extracted
from the results of Aguilar-Benitez and
Salicio\protect\cite{aguilar81}.}
\label{fig.fit3+4}
\end{figure}

\begin{figure}
\caption{The $K^{-}p$ scattering cross sections,
reaction cross sections and $\Sigma\pi$ mass spectrum calculated with
parameters of fits n7 and c1. Both fits include $S$, $P$, and $D$
waves, but were not adjusted to mass spectra data. Fit c1 used equal
weights for the data in the $\chi^{2}$ minimum search. The cross
sections data are same as in Fig.~\protect\ref{fig.fit3+4}.}
\label{fig.fitadd}
\end{figure}

\begin{figure}
\caption{The threshold  branching ratios for $K^{-}p$  to neutral
states $R_{n}$, to charged states $R_{c}$, and to charged $\Sigma\pi$
states $\gamma$. The comparisons in the top part of the figure
indicate that CBM fits adjusted to the $\Sigma\pi$ mass spectrum tend
not to agree with $\gamma$. The comparisons in the bottom indicate
that the original Schnick--Landau potential
parameters\protect\cite{schnick87} did not provide good agreement for
$\gamma$, but the update (solid curve) by Tanaka and
Suzuki\protect\cite{tanaka} does. The data are from Humphrey et
al.\protect\cite{humphrey62}, Tovee et al.\protect\cite{tovee71},
Nowak et al.\protect\cite{nowak78}, and Goossens et
al.\protect\cite{goossens80}.}
\label{fig.br}
\end{figure}

\begin{figure}
\caption{The $P_{13}$, $\pi\Lambda$ and $D_{03}$, $\bar{K}N$ scattering
amplitudes of fits 3 and 4. The $\mbox{Im } f$ in the lower halfs of
the figure peaks at the resonance energy for which the $\mbox{Re} f$
passes through zero in the upper part. The arrows show the $\Sigma\pi$
and $\bar{K}N$ thresholds.}
\label{fig.TPD}
\end{figure}

\begin{figure}
\caption{The $\bar{K}N$ $S_{01}$ scattering amplitudes for fits 1-4.
The arrows show the $\Sigma\pi$ and $\bar{K}N$ thresholds. The
resonance and threshold behaviors appear mixed.}
\label{fig.TS}
\end{figure}

\begin{figure}
\caption{The imaginary parts of the $\bar{K}N$ $S$--wave
scattering amplitudes of fit 2 (upper part) and fit 3 (lower part) as a
function of complex energy.  The projection onto the complex energy
plane shows the contours of the $T$ matrix. The tick mark to the right
of $1400$ on the real energy axis is the $\bar{K}N$ threshold energy.}
\label{fig.3D}
\end{figure}

\begin{figure}
\caption{The shift  and width  due to the strong interaction of the
$1S$ level in kaonic hydrogen. The data are from Davies et
al.\protect\cite{davies79}, Izycki et al.\protect\cite{izycki80}, and Bird et
al.\protect\cite{bird83}. The triangles
are prediction of the  different CBM fits.
All experiments indicate a positive $\epsilon$ while all CBM
calculations indicate a negative shift (to the less bound).}
\label{fig.H}
\end{figure}

\begin{table}
\begin{center}
\begin{tabular}{lccccc}
& 1 & 2 & 3 & 4 & 5 \\ \hline
Charge Basis  & $K^-p$ & $\bar{K}^0n$ & $\Sigma\pi$ $(I=0)$ &
$\Sigma\pi$ $(I=1)$ & $\Lambda\pi^0$ \\ \hline
Isospin Basis & $\bar{K}N (I=0)$ & $\bar{K}N (I=1)$ &
$\Sigma\pi$ $(I=0)$ & $\Sigma\pi$ $(I=1)$ & $\Lambda\pi^0$ \\
\end{tabular}
\end{center}
\caption{Channel assignment for charge and isospin bases.}
\label{tab.chnl}
\end{table}

\begin{table}
\begin{center}
\begin{tabular}{ccc|ccc|ccc}
 & $\lambda^\Lambda_\alpha$& $\lambda^\Sigma_\alpha$ & \multicolumn{3}{c|}
{($\lambda^{t,I=0}_{\beta\alpha},
\lambda^{s,I=0}_{\beta\alpha}$)} &
\multicolumn{3}{c} {($\lambda^{t,I=1}_{\beta\alpha},
\lambda^{s,I=1}_{\beta\alpha}$)} \\ \hline
 & & &$\bar{K}N$ & $\Sigma\pi$ & $\Lambda\pi$ & $\bar{K}N$ &
$\Sigma\pi$ & $\Lambda\pi$ \\ \hline
$\bar{K}N$ & $\sqrt{2}$ &
$\sqrt{\frac{8}{3}}$& $(-\frac{3}{2}, -\frac{3}{2})$
& $(-\frac{\sqrt{6}}{4}, \frac{1}{2\sqrt{6}})$ & $(0,0)$ & $(-\frac{1}{2},
\frac{1}{6})$ & $(-\frac{1}{2}, \frac{1}{6})$ & $(\frac{\sqrt{6}}{4},
\frac{\sqrt{6}}{4})$\\
$\Sigma\pi$ &$\sqrt{3}$ & $-\sqrt{\frac{8}{3}}$
& $(-\frac{\sqrt{6}}{4}, \frac{1}{2\sqrt{6}})$
& $(-2, -\frac{4}{3})$ & $(0, 0)$
            & $(-\frac{1}{2}, \frac{1}{6})$ & $(-1, -\frac{2}{3})$ &
$(0, -\frac{2}{\sqrt{6}})$ \\
$\Lambda\pi$ & $0$ & 2& $(0, 0)$ & $(0, 0)$ & $(0, 0)$
             & $(\frac{\sqrt{6}}{4},
\frac{\sqrt{6}}{4})$ & $(0, -\frac{2}{\sqrt{6}})$ & $(0, 0)$
\end{tabular}
\end{center}
\caption{The coupling constants $\lambda^\Lambda_\alpha$ and
$\lambda^\Sigma_\alpha$ for channel $\alpha$, and spacetime coupling
constants ($\lambda^{t,I}_{\beta\alpha},
\lambda^{s,I}_{\beta\alpha}$) for channels $\alpha$ and $\beta$.}
\label{tab.couple}
\end{table}

\begin{table}
\begin{center}
\begin{tabular}{lrrrrrrr}
Fit  & $S$ & $P,D$ & $M(\Sigma\pi)$ & $M(\Lambda\pi)$ & BR &
$\chi^{2}$\\ \hline
1 & $\surd$ & $\times$ & $\times$ & $\times$ & $\times$ & $\Delta \sigma$\\
2 &$\surd$ & $\times$ & $\surd$ & $\times$ & $\times$ &$\Delta\sigma$\\
Tri & $\surd$ & $\times$ & $\surd$ & $\times$ & $\times$ &$\Delta\sigma$\\
3 & $\surd$ & $\surd$ & $\surd$ & $\surd$ & $\surd$ &$\Delta\sigma$\\
4& $\surd$ & $\surd$ & $\surd$ &$\surd$ & $\times$ &m\\
n7 & $\surd$ & $\surd$ & $\times$ & $\times$ & $\times$ &$\Delta\sigma$\\
c1 & $\surd$ & $\surd$ & $\times$ & $\times$ & $\times$ &m\\
\end{tabular}
\end{center}
\caption{Characteristics of different fits.}
\label{tab.fits}
\end{table}

\begin{table}
\begin{center}
\begin{tabular}{lrrrrrrr}
Fit & $R$ & $M_{S01}$ & $M_{P13}$ &$M_{D03}$ & $f^{I=0}$ & $f_K^{I=1}
$ & $f_\pi^{I=1}$ \\ \hline 1 & 1.29 & 1554 & - & - & 87 & 80 & 75 \\
2 & 0.95 & 1588 & - & - & 126 & 82 & 146 \\ Tri & 1.00 & 1630 & - & -
& 120 & 100 & 110 \\ 3 & 1.23 & 1577 & 1421 & 1558 & 96 & 100 & 79 \\
4 & 1.32 & 1542 & 1417 & 1546 & 103 & 96 & 78 \\ add1 & 1.34 & 1564 &
-1249 & 1553 & 92 & 80 & 88 \\ add2 & 1.19& 1623& -12682 & 1568 & 90&
87& 80\\
\end{tabular}
\end{center}
\caption{Parameters of different fits; $R$ is in fm.,
others in MeV.}
\label{tab.params}
\end{table}

\begin{table}
\begin{center}
\begin{tabular}{rrrr}
Fit& $E(S)$ & $E(P)$& $E(D)$\\
\hline
1 & (2, -27), (-104, $-6 \cdot 10^{-5}$)& -&- \\ 2 & (17, -31), (-80,
-76)&- &-\\ Tri & (13, -12), (-95, -76) &- &-\\ $V_{r2}$\cite{fink} &
(-52, -39) & -&-\\ $V_{AHW}$\cite{alberg76} & (-14, -25) &- &-\\ 3 &
(5, -21), (-129, -4) &(-48, -12) &(81, -6)\\ 4 & (1.2, -13), (-114,
-74)& (-46, -12)&(83, -6)\\
\end{tabular}
\end{center}
\caption{The pole positions  in complex energy (in MeV)
for $S-D$ waves relative to the $\bar{K}N$ threshold at 1432 MeV.
Numerical uncertainty is $\sim \pm2$ MeV}.
\label{tab.pole}
\end{table}

\end{document}